\documentclass[aip,jcp,amsmath,amssymb,preprint]{revtex4-1}
\usepackage{graphicx, url, float}
\graphicspath{{Images/}}

\draft 

\begin{document}

\preprint{AIP}

\title{Efficiently Sampling Conformations and Pathways Using the Concurrent Adaptive Sampling (CAS) 
Algorithm}


\author{Surl-Hee Ahn}
\email{sahn1@stanford.edu}
\affiliation{Chemistry Department, Stanford University, Stanford, CA 94305}

\author{Jay W. Grate}
\email{jwgrate@pnnl.gov}
\affiliation{Pacific Northwest National Laboratory, Richland, WA 99352}

\author{Eric F. Darve}
\email{darve@stanford.edu}
\affiliation{Mechanical Engineering Department, Stanford University, Stanford, CA 94305}


\date{\today}

\begin{abstract}
Molecular dynamics (MD) simulations are useful in obtaining thermodynamic and kinetic properties of bio-molecules but are limited by the timescale barrier, i.e., we may be unable to efficiently obtain properties because we need to run microseconds or longer simulations using femtoseconds time steps. While there are several existing methods to overcome this timescale barrier and efficiently sample thermodynamic and/or kinetic properties, problems remain in regard to being able to sample unknown systems, deal with high-dimensional space of collective variables, and focus the computational effort on slow timescales. Hence, a new sampling method, called the ``Concurrent Adaptive Sampling (CAS) algorithm," has been developed to tackle these three issues and efficiently obtain conformations and pathways. The method is not constrained to use only one or two collective variables, unlike most reaction coordinate-dependent methods. Instead, it can use a large number of collective variables and uses macrostates (a partition of the collective variable space) to enhance the sampling. The exploration is done by running a large number of short simulations, and a clustering technique is used to accelerate the sampling. In this paper, we introduce the new methodology and show results from two-dimensional models and bio-molecules, such as penta-alanine and triazine polymer.
\end{abstract}

\pacs{Valid PACS appear here}
\maketitle 

\section{\label{sec:intro} Introduction}

Computational modeling of bio-molecules is an essential tool that gives us insight into mechanisms of bio-molecules that experiments fail to capture. We can easily simulate various bio-molecules of manageable size up to $\mu\mbox{s}$, which enables us to see realistic pathways and intermediates. However, it is still difficult to uncover most of the pathways and intermediates using all-atom simulations because of the temporal gap between simulation that requires a time step in femtosecond and biological systems with timescales of milliseconds. In addition, the simulated bio-molecules often stay trapped in metastable states, and no significant conformational change can be observed for a long time.

Consequently, several methods have been developed for all-atom molecular dynamics (MD) simulations to overcome these difficulties. One class of methods applies a biasing potential to force the system to move away from the metastable state, namely umbrella sampling, metadynamics, hyperdynamics, accelerated MD (aMD)\cite{torrie1977, laio2002, laio2008, barducci2008, bussi2006, voter1997, hamelberg2004, miao2015}, and adaptive biasing force (ABF)\cite{rodriguez-gomez2004, rodriquez-gomez2006, pohorille2006, darve2001, darve2002, darve2008, comer2015}. Umbrella sampling can sample specific regions of phase space by adding a restraining potential to the system's potential to keep the system close to those specific regions\cite{torrie1977}. Metadynamics, on the other hand, can quickly compute the free energy landscape by filling the visited places with ``Gaussians" or small repulsive Gaussian potentials, which forces the system to escape from local minima\cite{laio2002, laio2008, barducci2008, bussi2006}. Similarly, hyperdynamics and aMD can also quickly reconstruct the free energy landscape by raising the energy in low energy, metastable regions\cite{voter1997, hamelberg2004, miao2015}. Finally, ABF calculates the first derivative of the free energy landscape, which is used to bias the simulation to overcome large energy barriers more easily and improve sampling, and ensures the system to stay close to statistical equilibrium\cite{darve2008, comer2015}.

However, these methods are only able to have one or few collective variables, which is limiting in cases where many collective variables are used to characterize a bio-molecule's conformation. Since it is very challenging to find the few essential collective variables that can typically characterize the bio-molecule's conformation, in practice it is desirable to use many of them in the hope that the actual collective variable of interest is a function of the ones we selected. In addition, metadynamics requires the collective variables to be differentiable, which is also limiting in cases where we have non-differentiable collective variables, such as the number of hydrogen bonds.

Another class of methods changes the temperature of the system to sample states that are difficult to reach at the original temperature, namely replica exchange or parallel tempering and temperature-accelerated dynamics (TAD)\cite{sugita1999, voter2000}. Specifically, replica exchange enhances sampling by having copies or replicas of the same system at different temperatures and exchanging them periodically while maintaining an equilibrium canonical ensemble distribution for each temperature\cite{sugita1999}. Even though replica exchange maintains detailed balance for an extended ensemble of canonical states, it alters the actual kinetics of the system by exchanging states from different temperatures. Hence, we are unable to obtain real kinetic pathways from the method. TAD, on the other hand, raises the temperature and allows only those events that should occur at the original temperature to preserve correct dynamics\cite{voter2000}. However, the system is required to have minor anharmonic effects. 

An alternative method that can be used instead is building Markov state models (MSMs). In this method, the conformational space is divided into kinetically-relevant macrostates and a number of trajectories are run to compute transition probabilities, overall reaction rates, and other kinetic and thermodynamic properties\cite{pande_msmbook, bowman2009progress, bowman2009ge, pan2008, schutte2011}. This way, transition regions and long timescale events can be identified efficiently. 

However, the macrostate decomposition must be Markovian, which is not the case in most instances. As a rule of thumb, the lag time $\tau$, or the time between two macrostates, is chosen to be long enough, and the macrostates are chosen to be small enough so that transitions are Markovian\cite{bowman2009progress, pan2008}. However, controlling the Markovian error may be difficult or even practically impossible. One option is to use smaller macrostates such that the relaxation times inside a macrostate are very small. However, this leads to a significant increase in computational cost. 

Another option is to increase the lag time $\tau$. In practice though, it is very difficult to determine when the MSM is converging with respect to the lag time (see Suarez\cite{suarez2016} and the Supplemental Information). The rate is strictly determined by the eigenvalues of the infinitesimal generator of the stochastic system. As a result, the lag time $\tau$ effectively represents a discrete approximation of this generator. As $\tau$ increases, the mean first passage time will tend to increase, eventually exhibiting a linear dependency with $\tau$. Consequently, as the lag time increases, Markovian effects go down but errors due to the time discretization $\tau$ increase. In some cases, convergence can be very difficult to detect and may in effect never happen (see Supplemental Information\cite{suarez2016}).

To overcome the limitations that result from the Markovian assumption, the weighted ensemble (WE) method can be used instead\cite{huber1996, zhang2010, bhatt2010, zhang2007, suarez2014, suarez2016, abdul2012, costaouec2013, badi2014, izaguirre2015, trott2016}. Similar to MSMs, the WE method divides the conformational space into macrostates and runs a fixed number of short trajectories or ``walkers" with simulation length $\tau$ within each macrostate. The walkers carry probabilities or ``weights" that sum up to 1, and these weights eventually converge to steady-state weights. Unlike MSMs, however, there is no need to adjust the simulation time $\tau$ and the macrostate decomposition to control the accuracy. This is because the Markovian assumption is not required and hence, there is no inherent statistical bias upon convergence. The WE method yields unbiased and exact results in the absence of statistical errors (see Chapter 7 by Darve and Ryu in Schlick\cite{schlick2012}; also, Darve\cite{darve2013}). But because the Markovian assumption is not used, the WE method requires global convergence of the macrostate weights, which leads to a larger computational cost than MSMs (Chapter 7 in Schlick\cite{schlick2012}). Hence, since MSMs have uncontrollable errors unless the macrostates are chosen carefully, WE is preferable in many cases as it is more robust.

We now explain the WE method in more detail. The collective variables to keep track of (e.g., dihedral angles, bond distances) need to be chosen beforehand. Their values determine which macrostate each walker belongs to at each step. The macrostates form a partitioning of the collective variable space. Then, the walkers are run for $\tau$ amount of time and are binned in macrostates according to their new collective variables' values. Within each macrostate, a fixed target number of walkers or $n_w$ is maintained by merging or splitting walkers in a statistically correct way. This process is referred to as ``resampling." The reason for this is to maintain a constant stream of walkers going from one macrostate to another irrespective of the energy barrier height. If this was not the case, then the walkers would be depleted in macrostates near an energy barrier or overcrowded in macrostates at low energy, and they would not be able to overcome energy barriers and sample rare pathways and intermediates. Finally, these same steps are repeated until convergence.

However, the WE method loses efficiency if the macrostates are not correctly defined. In this case, sampling relevant regions for computing reaction rates takes longer and is less accurate\cite{aristoff2016}. For instance, if our partitioning is fine, then we end up having too many walkers to simulate. On the other hand, if our partitioning is coarse, then the walkers are unable to easily go over the energy barriers by having to go over energy barriers within the macrostate first, and trajectories of different walkers remain correlated for a long time. As a result, we end up with larger standard deviations. By itself, the WE method is not ideally suited to sample high dimensional spaces of collective variables because of the number of macrostates required to partition the collective variable space.

Considering these factors, we developed a method that aims to solve these aforementioned problems called the ``Concurrent Adaptive Sampling (CAS) algorithm." Similar to the WE method, the CAS algorithm runs a number of short simulations or ``walkers" for each macrostate and maintains a fixed target number of walkers for each macrostates so that macrostates are constantly sampled irrespective of their free energy barriers. Unlike the original implementations of the WE method, however, the CAS algorithm constructs macrostates based on (an approximation of) the committor function, which is the probability to reach the product state before reactant state from a given point. Each macrostate approximates an isocommittor surface. This guarantees that the walkers can make progress in sampling the reactant to product pathways of interest, while keeping the computational cost under check.

Using the exact committor function would lead to an optimal partitioning of phase space and the best possible WE sampling. However, this  problem in itself is as complicated or more than computing rates and pathways of interest. Consequently, we have to use an approximation of the committor function. This can be done in different ways. In this paper, we explore two options. First, the committor function can be computed adaptively as the simulation is on-going. We start from an initial guess, and refine it using simulation data as it is produced. Second, we run an initial brute force simulation for a short period to determine some of the pathways and approximate the committor function based on this partial sampling. An important observation is that obtaining an accurate estimation of the committor function is not required. Improvements in sampling and convergence can be observed even with a simple estimate of the committor function. In practice, we have observed that even short brute force simulations are sufficient to get a good partitioning and fast convergence of the CAS algorithm.

Besides knowing the collective variables and the reactant and product states beforehand, little a priori knowledge about the system is required. In addition, a macrostate is essentially a union of $n$-dimensional Voronoi cells, where $n$ equals the number of collective variables. A Voronoi cell is a region that is defined by its center and encompasses points that are closest to the center than any other center. The construction of Voronoi cells is detailed in Section~\ref{sec:macrostates}. Using $n$-dimensional Voronoi cells allows a feasible sampling of high dimensional spaces. The implementation is also relatively straightforward.

As previously mentioned, the committor function can either be computed during the simulation or before during a pre-processing step. This depends on whether we let the CAS algorithm adaptively construct macrostates as the simulation proceeds or use static macrostates throughout the simulation. If we use static macrostates, then we construct them to be isocommittor surfaces based on the (approximate) committor function. If we use adaptive macrostates, then we first let the Voronoi cells naturally follow the evolving probability distribution. In this case, the computational cost increases, and the number of constructed Voronoi cells initially grows. To mitigate this problem, we propose an algorithm to downsample these Voronoi cells in an optimal way, based on the sampled data up to this point. This is done by computing the transition matrix using the existing Voronoi cells. Then the committor function is approximated as the second left eigenvector. This is equivalent to approximating the exact committor function using piecewise constant basis functions (for which the functions are constant over each Voronoi cells). Then, we define new macrostates to be a union of Voronoi cells that have the same (similar) committor function values, so that each macrostate approximates an isocommittor surface. Within each macrostate, we resample so that we end up with a fixed number of walkers. Then, we let the Voronoi cells adaptively evolve again. 

This way, we control the computational cost by discarding walkers that are orthogonal to the pathway and keeping walkers that are progressing along the pathway. This guarantees an efficient sampling and progression along the reactant-to-product pathways. Note that our clustering method aims to cluster Voronoi cells based on their dynamic similarity rather than geometric or energetic similarity. Geometric clustering methods and Perron Cluster Cluster Analysis (PCCA) are sometimes used in order to build an MSM and to identify metastable states of the system, respectively\cite{pande_msmbook}. By clustering dynamically similar Voronoi cells via the committor function, we end up sampling dynamically important pathways and intermediates more efficiently.

The idea of adaptive macrostates has been discussed previously in the literature\cite{huber1996, zhang2010, bhatt2012, adelman2013, dickson2014}. In particular, the WE-based string method adaptively constructs macrostates that form the principal reaction pathway and achieves lower error and true mean first passage times more quickly than conventional methods\cite{adelman2013}. However, the method is poorly suited for finding multiple reactions and pathways and does not work when the reaction pathway is not well-defined. This is a common problem unless the system's temperature is low. The minimum free energy pathway can be easily defined at zero temperature. At higher temperatures, its definition is more ambiguous.

The WExplore method, on the other hand, adaptively constructs macrostates in a hierarchical fashion and effectively maps out the free energy landscape with pathways and intermediates that are not known a priori\cite{dickson2014, dickson_rna2014}. To control the computational cost, the WExplore method sets a strict limit on the total number of macrostates by pre-defining the following parameters: number of hierarchy levels, macrostate sizes for each level, and maximum number of branching macrostates for each level. Although the WExplore method is able to dynamically define regions and cover the entire landscape efficiently, the critical distances associated with each level needs to be defined carefully, and since it does not use the committor function, the method is heuristic\cite{dickson2014}. In this sense, the CAS algorithm is easier and more appropriate, since we only need to define the reactant and product states and calculate the committor function to have an optimal partitioning (based on the data collected so far).

\section{\label{sec:methods} Methods}
\subsection{\label{sec:resampling} Resampling}

\begin{figure}
\centering\includegraphics[width=0.75\textwidth]{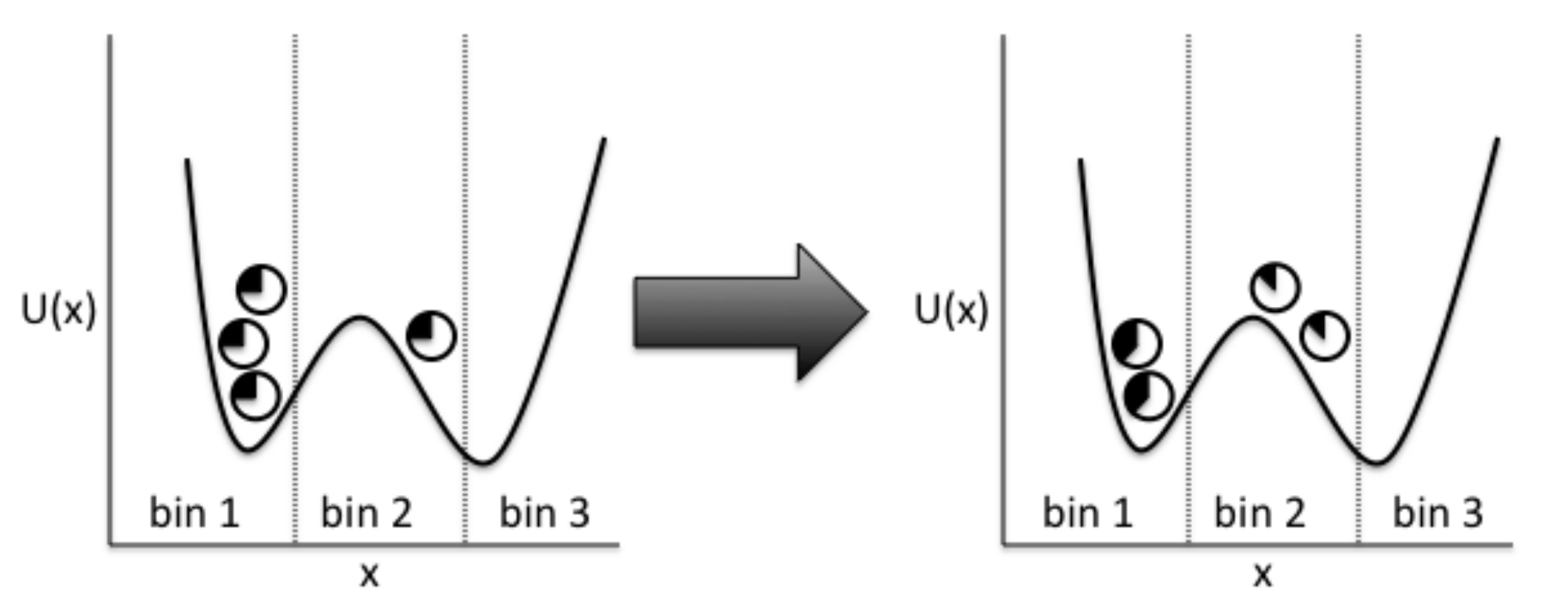}
\caption{\label{fig:resampling} Illustration of the resampling algorithm. Each macrostate or bin ends up with walkers with equal mean weight $W$. Here, one walker in Bin 1 is killed and its probability or weight is carried by the surviving walkers. A walker is created in Bin 2 by duplicating an existing walker and dividing its weight. After resampling, each bin has exactly the same number of walkers and weights.}
\end{figure}

In the CAS algorithm, after the walkers are run for $\tau$ amount of time and are binned to their corresponding macrostates according to their new collective variables' values, the walkers are resampled. That is, each walker $i$ gets assigned to the mean weight $W=\sum_i w_i/n_w$ in its respective macrostate, and $n_w$, or target number of walkers per macrostate, walkers are maintained in each macrostate. This resampling algorithm is illustrated in Fig.~\ref{fig:resampling} and was first suggested by Darve and Izaguirre (Chapter 7 in Schlick\cite{schlick2012}). This is different from the WE method, which generates walkers with weights between $W/2$ and $2W$ but not exactly equal to $W$. However, assigning a constant weight has been proven to be optimal in terms of minimizing variance and statistical errors (Chapter 7 in Schlick\cite{schlick2012}). To illustrate how the mean weight $W$ is assigned to each walker, we first start out with a list of walkers to process for each macrostate. Then we sort the walkers in descending order based on their weights. The sorting of the weights helps reduce data correlation, since when a walker is split, the newly made walkers become correlated for some time. Then we encounter the following two cases for the walkers:

\begin{enumerate}
\item
$w_i \geq W$: In this case, the walker is split into an integer number of walkers of weight $W$, and the remainder is reinserted into the list of walkers to process.
\item
$w_i < W$: In this case, the walker is merged with other walkers in a statistically exact way to create a walker with a weight $\geq W$. For instance, if we have a walker with a weight $w_1$ and another walker with a weight $w_2$, we randomly select one of the walkers with probabilities $w_1/(w_1+w_2)$ and $w_2/(w_1+w_2)$ respectively and assign the chosen walker with a weight $w_1+w_2$ and so on. From this procedure, we can clearly see that each walker will end up with a weight $W$, and any number of walkers can be combined or split in this fashion without violating any rules of probability.
\end{enumerate}

In the end, we end up with $n_w$ walkers in each macrostate with equal mean weight of $W$. As mentioned previously, this is a crucial aspect of the WE method and the CAS algorithm since resampling allows rare regions to be sampled efficiently and continuously throughout the simulation.

\subsection{\label{sec:macrostates} Defining macrostates} 
As mentioned in Section~\ref{sec:intro}, the macrostates form a partitioning of the collective variable space and are essentially unions of $n$-dimensional Voronoi cells, where $n$ equals the number of collective variables. We can either adaptively construct or pre-define and fix the macrostates throughout the simulation. In the case of pre-defining and fixing macrostates, Section~\ref{sec:spectral_clustering} will discuss in detail how to calculate the committor function, which is used to form the macrostates into isocommittor surfaces.

In the case of adaptively constructing macrostates, we need to use $n$-dimensional spheres of radius $r$, which are used to define the newly created Voronoi cells' centers. Before the Voronoi cells are constructed, all of the walkers have ran for $\tau$ amount of time and are ready to be binned to their corresponding Voronoi cells. The basic outline of how the Voronoi cells are created during a single simulation step is as follows:

\begin{enumerate}
\item The very first walker in the list of walkers to bin is binned to its own center. That is, a new center is created and is equal to the walker's collective variables' values.
\item For the subsequent walkers, they are tested to see if the distance between the walker's collective variables' values and any of the existing centers is less than or equal to $r$. If not, then a new center is created that is equal to the walker's collective variables' values.
\item If there is more than one center that is $r$ or less away from the walker, then the walker is binned to the center that is closest to the walker. 
\item After all of the walkers are binned to a center, then we go through all of the walkers once more to bin them to their true closest centers, since the centers have been created for one walker at a time.
\item We delete the centers that have no walkers, and the remaining centers become the Voronoi centers.
\end{enumerate}

This adaptive construction of Voronoi cells allows us to sample unknown systems without having to partition its collective variable space beforehand, which is especially suitable for high-dimensional systems. Note that the radius $r$ is chosen such that we have relatively fast relaxation times within the resulting Voronoi cells. Hence, $r$ should not be too big so that we have significant energy barriers within the resulting Voronoi cells. As a rule of thumb, small Voronoi cells are suitable for systems with high energy barriers at transition regions, whereas large Voronoi cells are suitable diffusive systems with low energy barriers.

Since the number of Voronoi cells and associated walkers can quickly increase after the simulation has proceeded for a number of steps, we need a way to control the number of walkers to make the CAS algorithm computationally tractable. The committor function is used to do that and will be explained in Section~\ref{sec:spectral_clustering}.

\subsection{\label{sec:spectral_clustering} Spectral clustering}
As was mentioned in Section~\ref{sec:macrostates}, the number of Voronoi cells and associated walkers may quickly increase as the bio-molecule's collective variable space is explored. In this case, the computational cost of running all of these walkers will be high, and sampling along the reactant to product pathway of interest will take much longer time. Hence, we need to calculate the committor function in order to resolve this issue. The committor function describes the probability to reach the product state before reaching the reactant state first. If the committor function of a macrostate is 0, then the macrostate is a reactant state, and if the committor function of a macrostate is 1, then the macrostate is a product state. In other words, we can characterize how close the macrostate is in terms of reaching the product state and exactly where the macrostate is in the reactant to product pathway of interest. 

In order to calculate the committor function, we first have to compute the transition matrix of the existing Voronoi cells for a number of steps. This is because we use the eigenvectors of the transition matrix, which are shown to be equivalent to computing an approximation of the committor function where the approximation is constant over each Voronoi cell, i.e., a piecewise constant approximation\cite{prinz_efficient2011}. Specifically, we approximate the committor function $\psi(x)$ to be:
\begin{equation}
\psi(x) \approx \rho_2(x)/\rho(x).
\label{eq:committor}
\end{equation}

Here, $\rho_2(x)$ denotes the eigenvector corresponding to the second largest eigenvalue $\lambda_2$, which represent probability changes of Voronoi cells in the pathway of interest. $\rho(x)$ denotes the eigenvector corresponding to the largest eigenvalue $\lambda_1 = 1$, which represent equilibrium weights of the Voronoi cells. Note that we have not normalized $\psi(x)$ so that it ranges from 0 to 1, but it is not necessary since this only shifts the values. With the committor function calculated, we can cluster the Voronoi cells by their committor function values. The union of Voronoi cells then becomes our new macrostates. Within each new macrostate, we can resample walkers and end up evenly covering the reactant to product pathway of interest with walkers and reducing walkers that are redundant or orthogonal to the pathway. Hence, we naturally call this method spectral clustering. We note that our spectral clustering is slightly different from the ones in published works, which use the first $k$ generalized eigenvectors to cluster a high-dimensional data into $k$ clusters using k-means\cite{ng2002, vonluxburg2007}. We only use the first two eigenvectors, and the number of clusters is set separately.

\begin{figure}
\centering\includegraphics[width=0.8\textwidth]{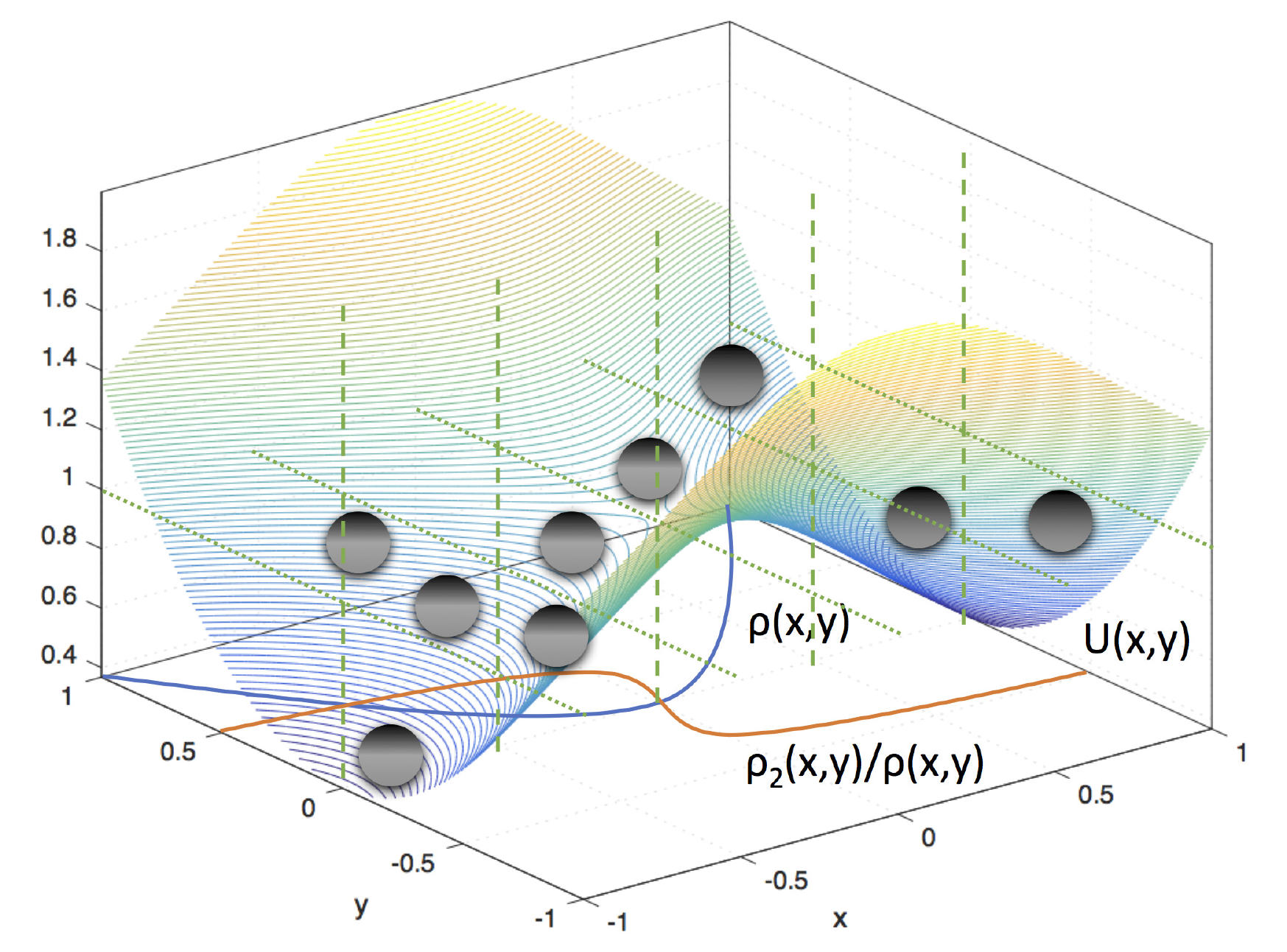}
\caption{\label{fig:spectral_clustering} Illustration of spectral clustering. Along with the free energy landscape $U(x,y)$, the equilibrium eigenvector $\rho(x,y)$ and the committor function $\rho_2(x,y)/\rho(x,y)$ are indicated. The walkers are represented as black circles. The partitioning from spectral clustering is also indicated by the vertical dashed lines. The horizontal dashed lines are present to elucidate which macrostates are orthogonal to the reactant to product pathway of interest. This way, the pathway gets evenly sampled and covered while reducing walkers that are orthogonal to the pathway.}
\end{figure}

To illustrate the method more clearly, we present the basic scheme of spectral clustering in Fig.~\ref{fig:spectral_clustering}. Here, $\rho(x,y)$ is indicated, and the equilibrium weights are highest in the two metastable states, which are the reactant and product states, respectively. In addition, $\rho_2(x,y)/\rho(x,y)$ is indicated, and the dynamics at this longest time scale is characterized by a global shift in probability density between the two states, which makes sense because the equilibration between the two would take the longest than any other non-stationary process in the system. Finally, with spectral clustering, the reactant to product pathway of interest is partitioned into dynamically distinct regions characterized by the values of $\rho_2(x,y)/\rho(x,y)$ and are marked with vertical dashed lines. This way, we increase sampling along the pathway while reducing sampling orthogonal to the pathway. After clustering and resampling, the walkers, which are represented as black circles, will evenly cover the pathway with new walkers in deficient regions and fewer walkers in oversampled regions.

The key idea is that we use importance sampling in spaces orthogonal to the reactant to product pathway of interest. Although pathways are not represented explicitly in our method, the use of the second eigenvector allows us to use a fine discretization along the pathway, while importance sampling is used in the orthogonal directions to make sure we control the number of Voronoi cells throughout the simulation. The second eigenvector is initially computed using incomplete information. However, the information provided is sufficient to control the number of Voronoi cells constructed and make sure it remains bounded, while allowing the system to make progress along the pathway. Ref.~\onlinecite{mcgibbon2016} also states that the natural reaction coordinate is essentially the second eigenvector.  

Up to this point, the committor function has only been used to cluster the adaptively created Voronoi cells to newly define macrostates throughout the simulation. Alternatively, we can use the committor function to initially partition the collective variable space into macrostates and run the CAS algorithm with these fixed, static macrostates.  The macrostates use an approximate committor function, but if we use the exact committor function, then this static partitioning can be proven to be the most optimal partitioning such that the CAS algorithm converges after one resampling step in the limit of having infinite number of walkers per macrostate or $n_w = \infty$. That is, we do not need to relax the walkers' weights to steady-state to get the correct fluxes. This choice is also optimal because the accuracy of the flux becomes independent of the relaxation times for the walkers inside each macrostate. If the macrostates are not chosen correctly, the relaxation time inside macrostates will lead to long correlation times in the flux values, resulting in larger statistical errors. With an optimal choice, the relaxation times have no effect on the standard deviation of the fluxes. 

Since each macrostate has a constant committor function value, all of the walkers from a particular macrostate have the same probability of ending up in another macrostate. That is, the walkers do not need to relax or go over energy barriers within each macrostate to reach the correct flux, since they are enforced to go to their correct macrostates in the next step by being in isocommittor surfaces. Just as isocommittor surfaces are proven to be optimal milestones for milestoning, we now prove that isocommittor surfaces are optimal macrostates for the CAS algorithm\cite{vanden2008}.

\bigskip

\textbf{Proposition.} Let $A$ and $B$ be two metastable regions of interest and $\Omega$ denote the entire collective variable space.

The following assumptions are made:
\begin{enumerate}
\item The system's dynamics obeys detailed balance or time reversibility.
\item Macrostates are constructed such that the committor function value is constant in each macrostate (i.e., they are isocommittor surfaces).	
\item $n_w = \infty$.
\end{enumerate}

The result is then all of the walkers in each macrostate, regardless of the walkers' distribution and positions, have the same probability to end up in another macrostate. Said otherwise, it is not necessary to relax the distribution of walkers inside each macrostate to obtain the correct fluxes between A and B and vice versa. Hence, we obtain exact rates going from $A$ to $B$ and vice versa in one step.

\textbf{Proof.}  First, let $q(x)$ denote the forward commitor function or the solution of the following Dirichlet partial differential equation problem with respect to $L$, which denotes the infinitesimal generator of the diffusion that governs the system's dynamics.
\begin{equation*}
(Lq)(x) = 0\ \mbox{if}\ x\in\Omega\setminus (A\cup B)
\end{equation*}
\begin{equation}
q(x) = 0\ \mbox{if}\ x\in A
\label{eq:pde}
\end{equation}
\begin{equation*}
q(x) = 1\ \mbox{if}\ x\in B
\end{equation*}

The solution of Eq.~(\ref{eq:pde}) is the committor function that describes the probability to reach product state $B$ before reactant state $A$. Additionally, $1-q(x)$ denotes the backward committor function.

Now, let $N$ be the number of macrostates that divide up $q(x)$ and $z_1 = 0 < z_2 < z_3 < \cdots < z_N = 1$ denote uniformly spaced committor function values. Also, let $m_{ij},\ j = 1,...,n_i$ denote a Voronoi cell that has a center $x_{ij}$ and a committor function value of $z_i$, and $n_i$ be the total number of those Voronoi cells. Then we can construct the macrostates to be the following isocommittor surfaces or union of Voronoi cells that have the same committor function value:
\begin{equation}
M_i = \bigcup_{j = 1}^{n_i} m_{ij} = \bigcup_{j = 1}^{n_i} \{x\ |\ |x-x_{ij}| \leq |x-x_{kl}|, \ ij \neq kl, \ q(x) = z_i\}, \ i = 1,...,N.
\label{eq:isocommittor}
\end{equation}	

Hence, $A$ can be re-labeled as $M_1$ and $B$ as $M_N$. From Eq.~(\ref{eq:isocommittor}), we can expect the macrostates close to $M_1$ to have $0 < q(x) \ll 1$ and close to $M_N$ to have $0 \ll q(x) < 1$. Now, we will prove that all of the walkers in $M_i$ have the same probability to end up in another macrostate $M_j$, regardless of where the walkers are located within $M_i$. Note that this is typically not the case because where the walker ends up in the next step depends on its position within the current macrostate.

To see this, let the reactant state be the union of all of the states left of $M_i$, i.e., $\tilde{M_1} = \bigcup_{i = 1}^{i-1} \bigcup_{j = 1}^{n_i} m_{ij}$, a union of macrostates that have a committor function value $\leq z_{i-1}$, and the product state be the union of all of the states right of $M_i$, i.e., $\tilde{M_N} = \bigcup_{i = i+1}^{N} \bigcup_{j = 1}^{n_i} m_{ij}$, a union of macrostates that have a committor function value $\geq z_{i+1}$, and consider this new reaction. Then $M_i$ becomes an isocommittor surface for this reaction as well, and the committor function for this reaction becomes $\tilde{q}(x) = (q(x) - z_{i-1})/(z_{i+1}-z_{i-1}) = (z_i - z_{i-1})/(z_{i+1}-z_{i-1})$. Indeed, $\tilde{q}(x)$ still satisfies Eq.~(\ref{eq:pde}) and boundary conditions, since $\tilde{q}(x) = 0$ if $x\in \tilde{M_1}$ and $\tilde{q}(x) = 1$ if $x\in \tilde{M_N}$. Returning to our original setup, we get the following probabilities $p_{ik}$ to go to $M_k$ after being in $M_i$
\begin{equation}
p_{ik} = \begin{cases}
\frac{z_i - z_{i-1}}{z_{i+1}-z_{i-1}} & \mbox{if $k = i+1$, $i = 2,...,N-1$}\\
1-\frac{z_i - z_{i-1}}{z_{i+1}-z_{i-1}} = \frac{z_{i+1} - z_i}{z_{i+1}-z_{i-1}} & \mbox{if $k = i-1$, $i = 2,...,N-1$}\\
1 & \mbox{if $i=1, k=2$ or $i=N$, $k=N-1$}\\
0 & \mbox{otherwise}.
\end{cases}
\label{eq:probabilities}
\end{equation}

Taken together, all of the walkers in $M_i$ reach $M_k$ with the same probability $p_{ik}$, regardless of the walkers' distribution, since each $M_i$ is an isocommittor surface. Since no systematic errors are present and statistical errors become zero in the limit of $n_w = \infty$, we obtain the exact rates going from $A$ to $B$ and vice versa in one step in the limit of $n_w = \infty$. $\square$

\bigskip

Now that we have proven that using the committor function to create macrostates is the optimal choice, the specific steps taken for spectral clustering are listed below:

\begin{enumerate}
\item After the simulation has a number of Voronoi cells that is equal or greater than the pre-defined threshold number of Voronoi cells,  a transition matrix $T$ of existing Voronoi cells is calculated for a number of simulation steps. The Voronoi cells are fixed during these simulation steps so that $T$ can be calculated. Each entry $T_{ij}$ represents the weight that transitioned from Voronoi cell $i$ to Voronoi cell $j$ in one step. This can be done since each walker keeps track of its weight, previous coordinates, and current coordinates. However, since a transition matrix needs to fulfill detailed balance when a canonical ensemble simulation is run under Hamiltonian dynamics, each entry $T_{ij}$ is computed using the equations in Ref.~\onlinecite{prinz2011}, which are:
\begin{eqnarray}
C_{ij} & = & \frac{B_{ij}+B_{ji}}{2s} = C_{ji}\\
T_{ij} & = & \frac{C_{ij}}{\sum_{k=1}^{n}C_{ik}}.
\label{eq:trans_mat}
\end{eqnarray}

Here, $B_{ij}$ denotes the sum of the weights that went from Voronoi cell $i$ to Voronoi cell $j$ during $s$ number of steps, and $C_{ij}$ denotes the state-to-state time-correlation estimator that fulfills detailed balance. Using $C_{ij}$, we can calculate $T_{ij}$ using Eq.~(\ref{eq:trans_mat}). The detailed balance requirement also reduces the uncertainties of the committor function\cite{metzner2009}. This transition matrix $T$ corresponds to the ``graph Laplacian matrix" $L$ described in Ref.~\onlinecite{ng2002, vonluxburg2007}.

\item Then an eigendecomposition on $T$ is performed. From the properties of a transition matrix, the eigenvector corresponding to the largest eigenvalue $\lambda_1 = 1$ or $\rho(x)$ represents the equilibrium eigenvector, as previously mentioned. The rest of the eigenvectors correspond to non-stationary processes that are slower for eigenvectors corresponding to eigenvalues close to 1.

\item Using the equilibrium eigenvector, the eigenvector corresponding to the second largest eigenvalue $\lambda_2$ or $\rho_2(x)$ is normalized to be a good approximation of the committor function, as previously stated in Eq.~(\ref{eq:committor}). The normalized eigenvector entries represent probabilities of going from reactant to product and are the right quantities to use for clustering, since sampling along the pathway will increase while preserving the dynamics of the system. Hence, $\rho_2(x)/\rho(x)$ is used to cluster the Voronoi cells using k-means, and the resulting union of Voronoi cells become the new macrostates. The number of clusters is set beforehand.

\item In resampling the new macrostates, the target number of walkers per cluster or macrostate $n_{wc}$ is set so that the total number of walkers will be reduced overall. After resampling, the most probable Voronoi cells end up being mostly populated within the new macrostates.
\end{enumerate}

After spectral clustering is performed as listed above, normal CAS algorithm steps are taken until the number of Voronoi cells hits a certain threshold, which signals spectral clustering to be performed again. This method will be illustrated more clearly in Section~\ref{sec:examples}, where specific examples will be discussed.

\section{\label{sec:examples} Examples}
\subsection{\label{sec:toymodel} Two-dimensional system with one minimum energy pathway}
\begin{figure}
\centering\includegraphics[width=0.8\textwidth]{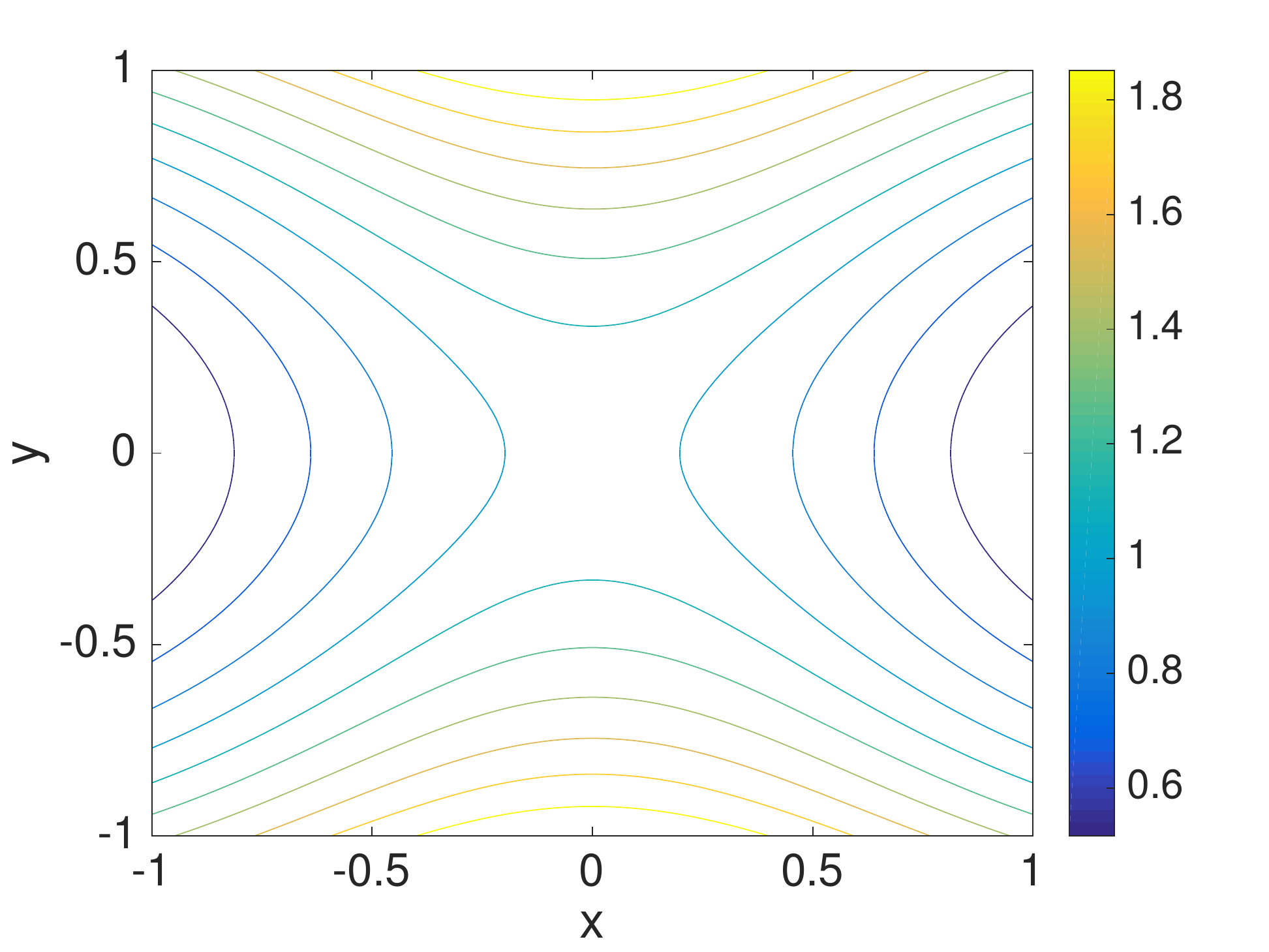}
\caption{\label{fig:toymodel} Two-dimensional potential surface described by $U(x,y) = e^{-x^2}+y^2$. The color bar indicates energy values.}
\end{figure}

For the simplest case, the CAS algorithm is tested on a two-dimensional potential surface described by $U(x,y) = e^{-x^2}+y^2$ taken from Ref.~\onlinecite{schlick2012}. The potential surface is bounded by $-10.0\leq x\leq 10.0$ and $-4.0\leq y\leq 4.0$. The Metropolis algorithm is used to move the walkers with $\Delta x = 0.05$ or $\Delta y = 0.05$ once per simulation step, and the target number of walkers per Voronoi cell $n_w$ is set to 100 for all of the simulations. As seen in Fig.~\ref{fig:toymodel}, the minimum energy basins are $A = \{(x,y) : -1.0\leq x\leq 0.0,\ -1.0\leq y\leq 1.0,\ \sqrt{(x+1)^2+y^2} \leq 0.4\}$ and $B = \{(x,y) : 0.0\leq x\leq 1.0,\ -1.0\leq y\leq 1.0,\ \sqrt{(x-1)^2+y^2} \leq 0.4\}$, and the minimum energy pathway is a straight pathway linking these two basins. The inverse temperature $\beta$ is set to 25.0, and half of the walkers initially start from $(-0.6, 0.0)$, or $A$, and the other half start from $(0.6, 0.0)$, or $B$. Since the initialization of walkers impacts the convergence time of the rates, the initial conditions are picked such that the simulations will reach convergence quickly. The forward (from $A$ to $B$) and backward (from $B$ to $A$) rates are measured by labeling walkers from $A$ and $B$ with colors, changing colors when walkers from $A$ reach $B$ and vice versa, and resampling each color within each Voronoi cell, as done in Ref.~\onlinecite{schlick2012}.

All of the simulations have the same fixed running time to ensure that the computational cost is the same across all simulations. The simulations differ from each other by having different radii $r$ and/or having spectral clustering turned on or not. These different simulation conditions are chosen to demonstrate the power of the CAS algorithm with spectral clustering compared with ``conventional" methods without spectral clustering. The committor function is used for spectral clustering to efficiently sample the slowest process, which in this case is the equilibration between the two basins $A$ and $B$. Three simulation runs are done for each kind of simulation and the standard deviation of the three runs is multiplied by 2, which approximately represent 95\% confidence interval, for error bars. 

\begin{figure}
\centering
\begin{tabular}{cc}
\includegraphics[width=80mm]{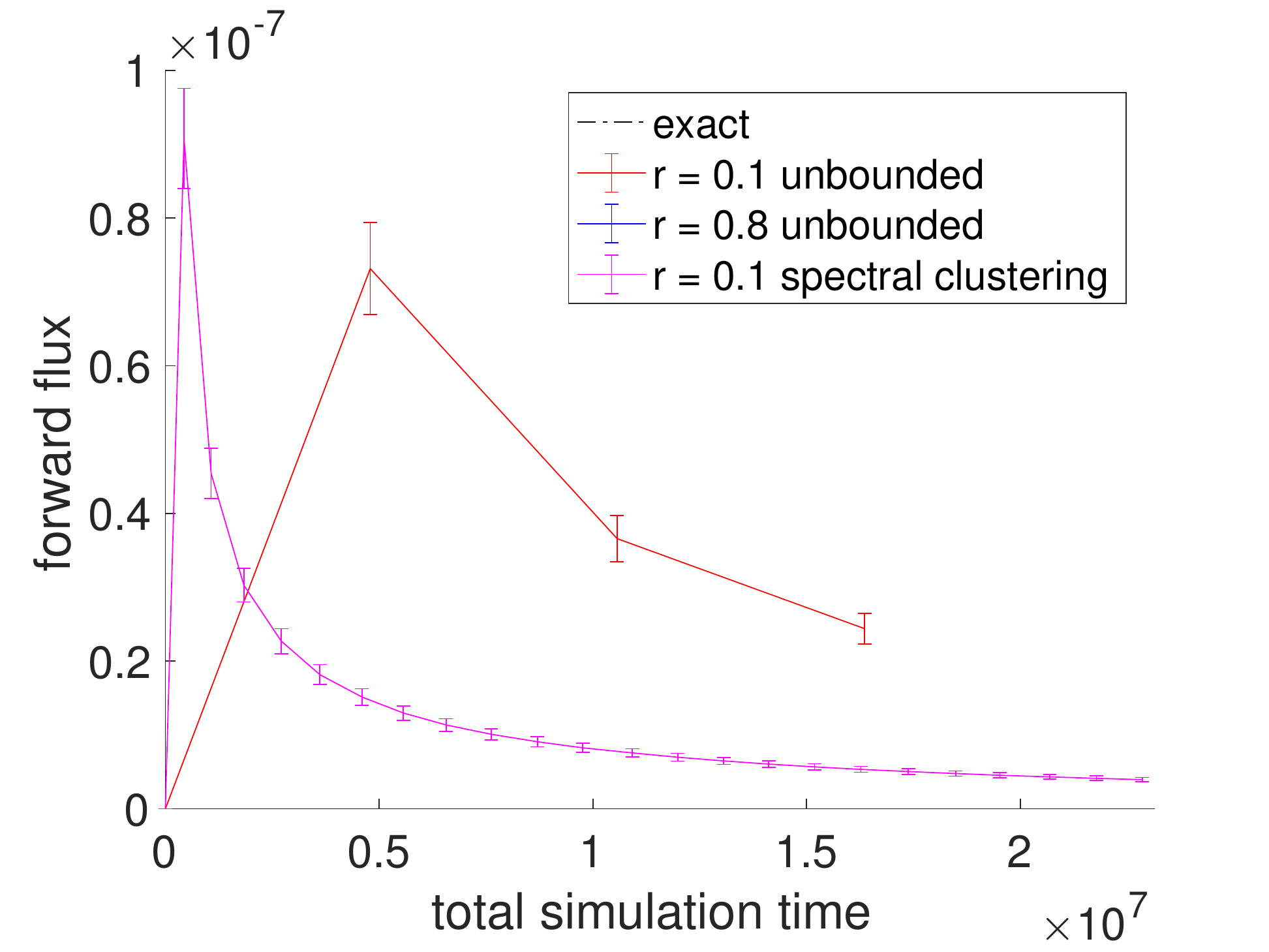} & \includegraphics[width=80mm]{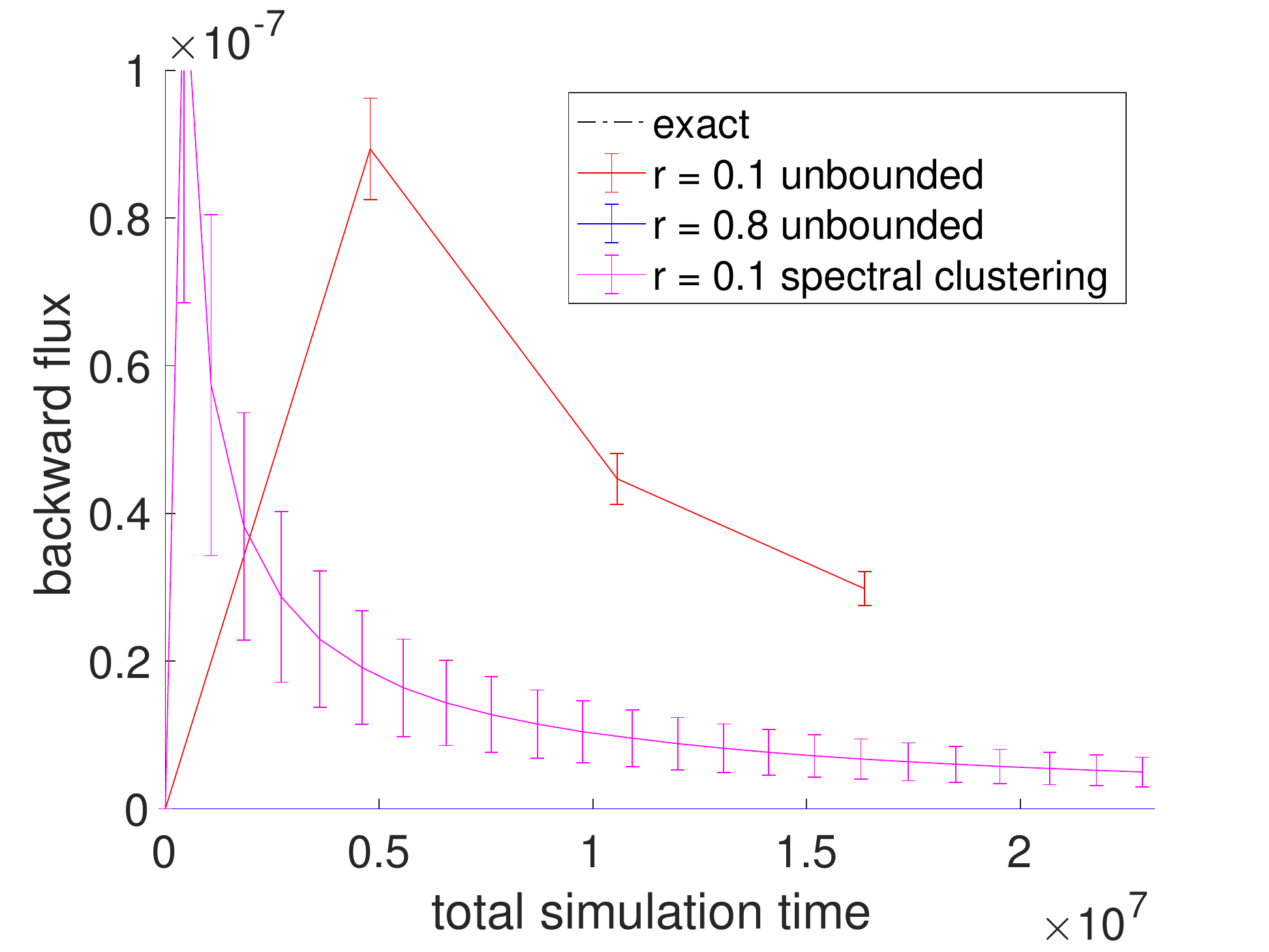} \\
(a) Forward rate ($A$ to $B$). & (b) Backward rate ($A$ to $B$). \\[6pt]
\end{tabular}
\caption{\label{fig:toymodel_fluxes} Rate comparisons among CAS algorithm simulations with and without spectral clustering. Other spectral clustering simulations with different parameters give similar results (data not shown).}
\end{figure}

To check whether the CAS algorithm with spectral clustering is more efficient than ``conventional" methods without spectral clustering, we plotted and compared the forward and backward rates between $A$ and $B$. Indeed as seen in Fig.~\ref{fig:toymodel_fluxes}, the CAS algorithm with spectral clustering is the one that is closest to converging to the correct rates within the fixed running time. When the Voronoi cells are chosen to be big ($r = 0.8$), they are unable to go over the energy barrier separating $A$ and $B$ and have walkers equilibrate between $A$ and $B$. In contrast, when the Voronoi cells are chosen to be small ($r = 0.1$), the number of Voronoi cells grows rapidly, and we end up wasting our efforts in covering every region of the potential surface. Since the potential surface is bounded, the number of Voronoi cells does not grow beyond a certain point. Nonetheless, the computational cost becomes large when many Voronoi cells are created, and spectral clustering proves to be useful in this scenario. With spectral clustering, we are able to focus our efforts in having walkers equilibrate between $A$ and $B$ and sample this slowest pathway and its intermediates. 

\begin{figure}
\centering
\begin{tabular}{cc}
\includegraphics[width=65mm]{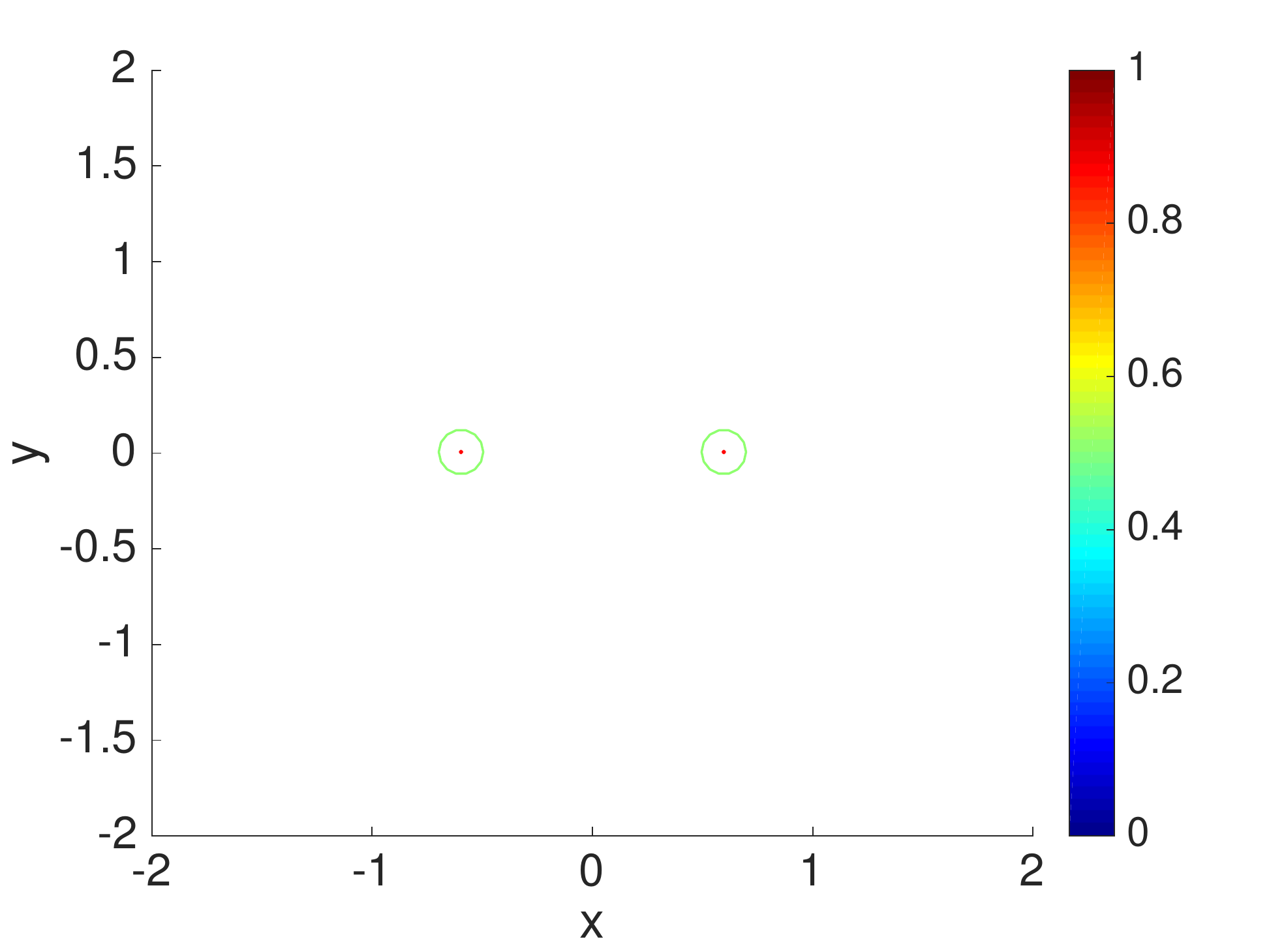} & \includegraphics[width=65mm]{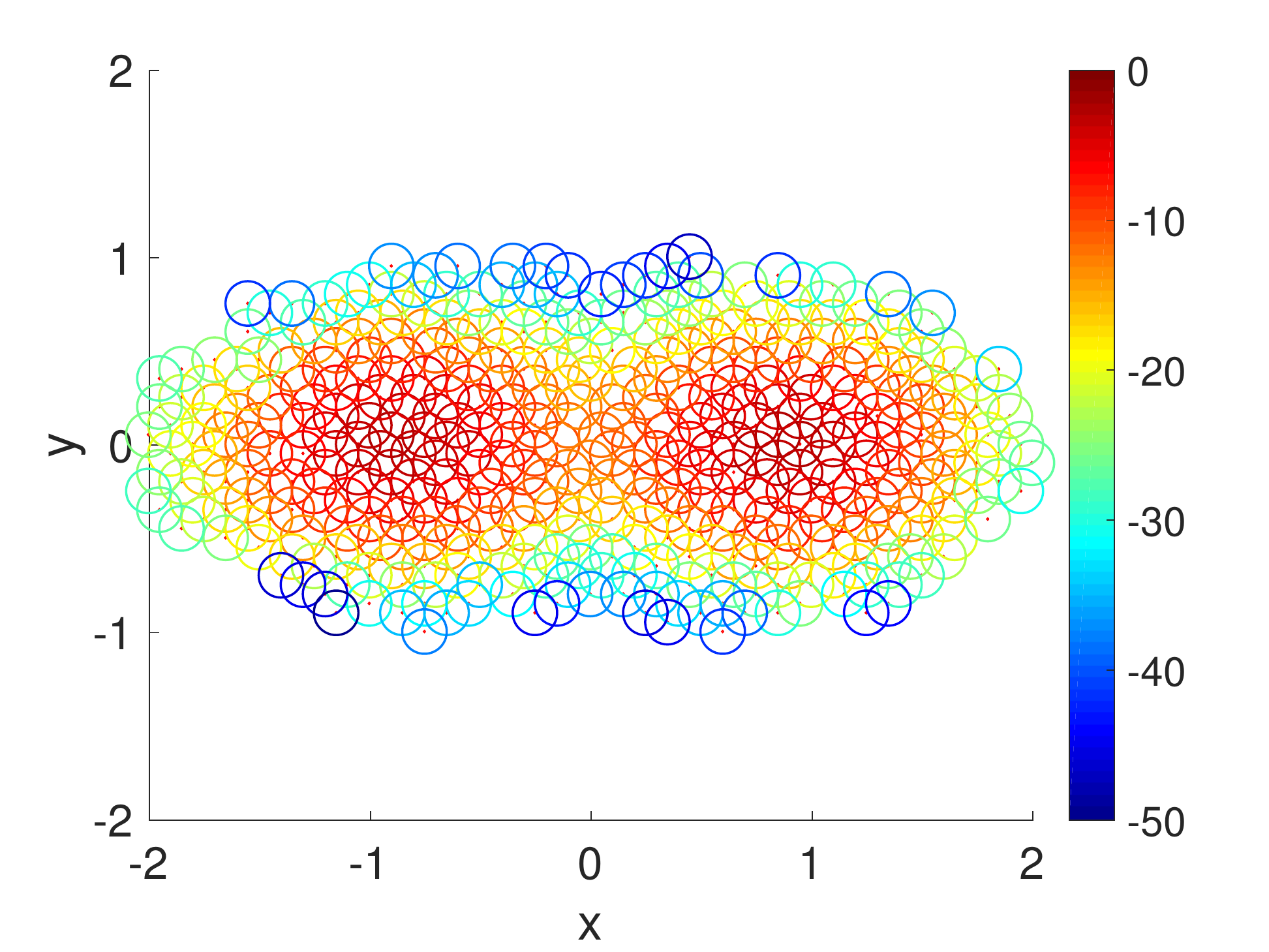} \\
(a) Initial condition. Step 0. & (b) Transition matrix calculation starts. Step 33.\\[6pt]
\includegraphics[width=65mm]{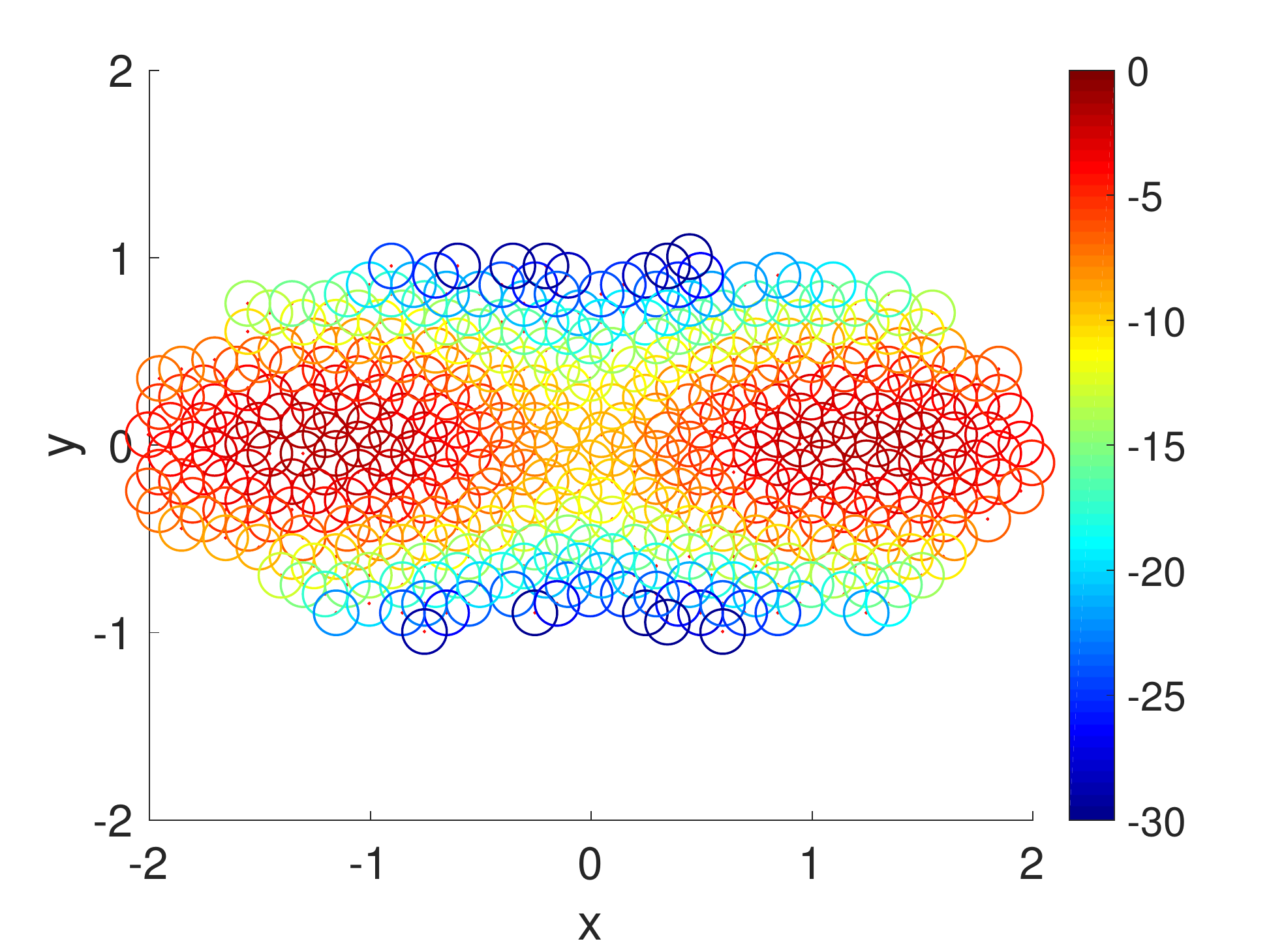} & \includegraphics[width=65mm]{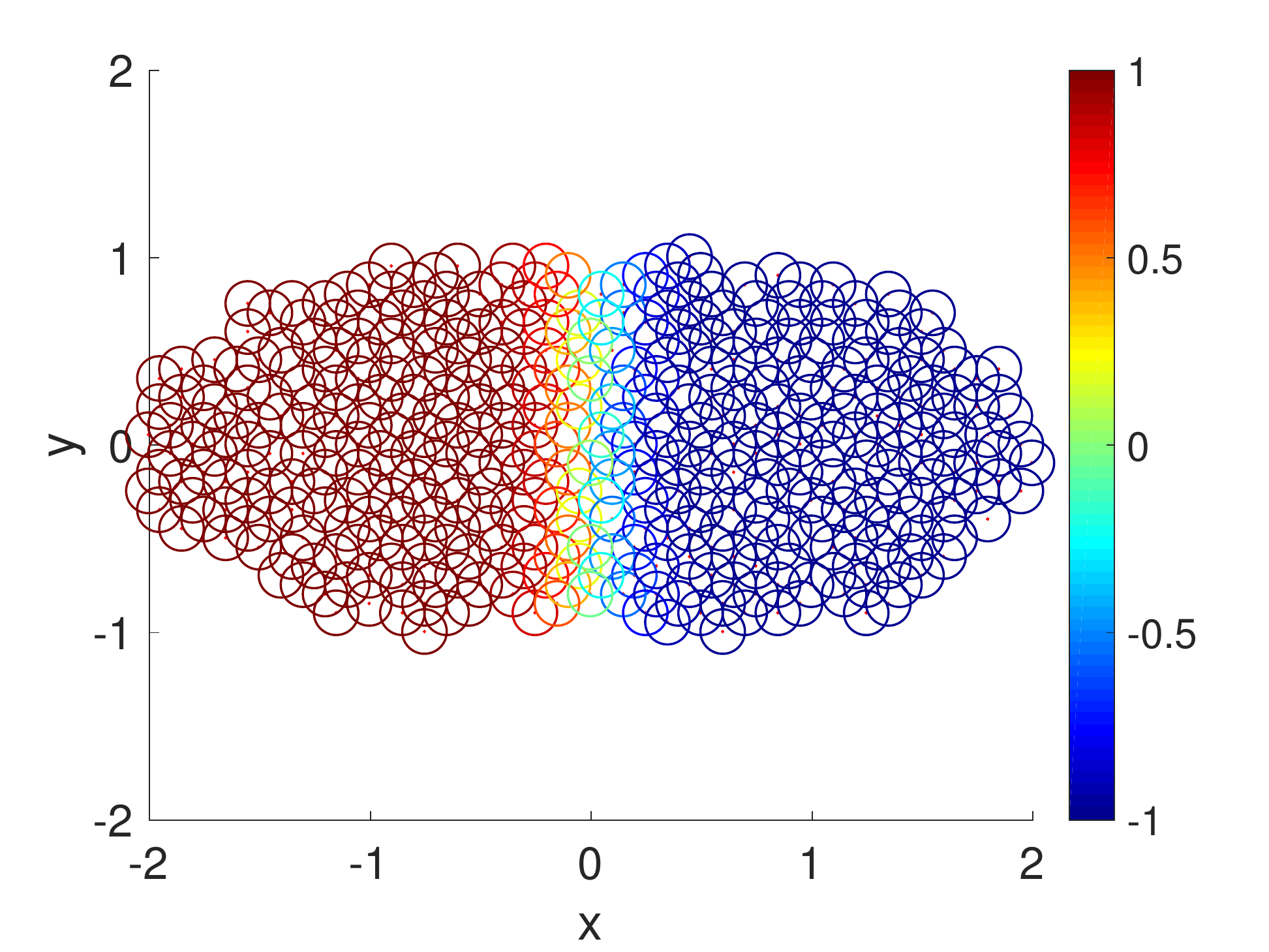} \\
(c) Equilibrium weights. Step 133. & (d) Committor function. Step 133. \\[6pt]
\includegraphics[width=65mm]{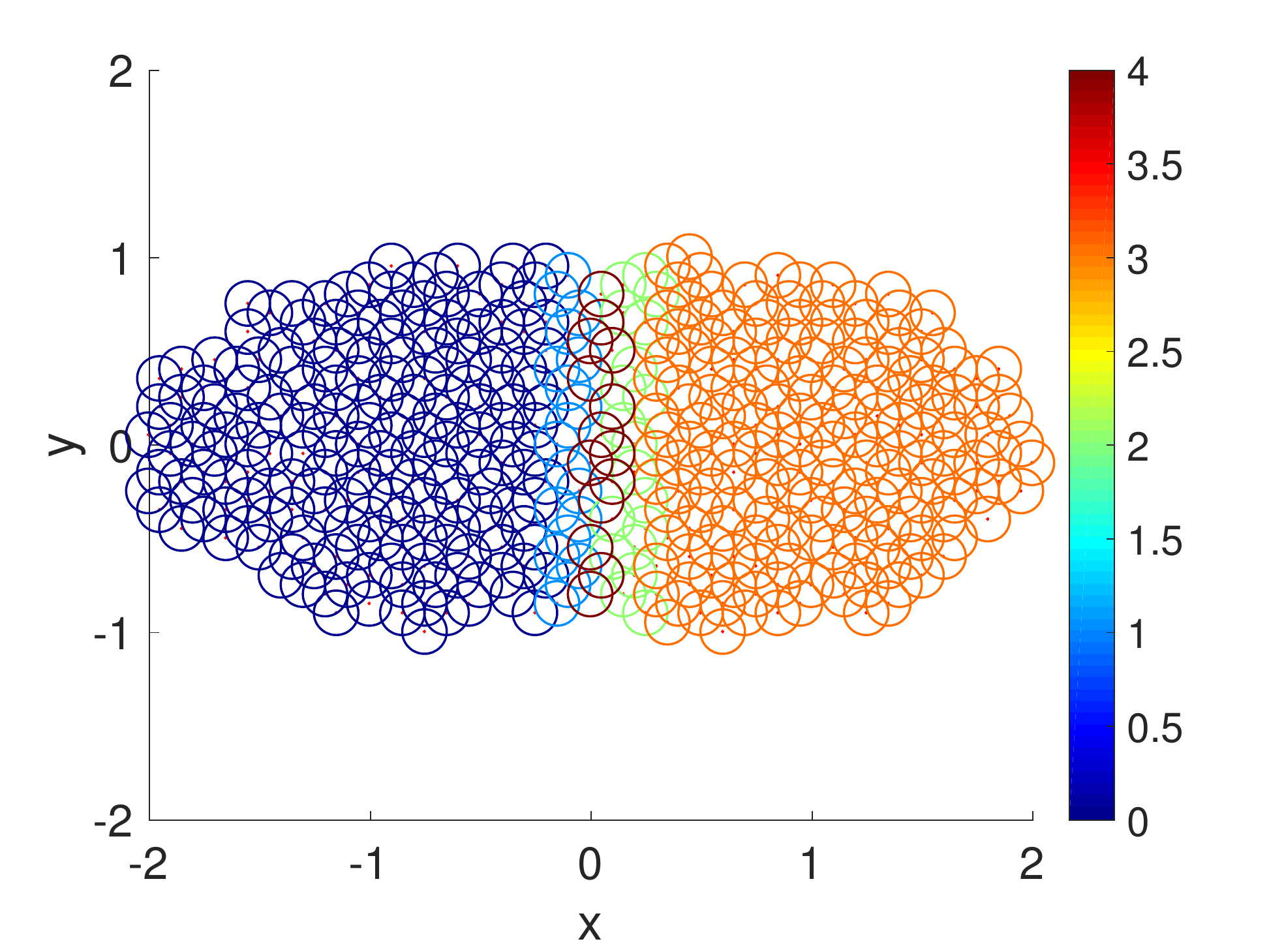} & \includegraphics[width=65mm]{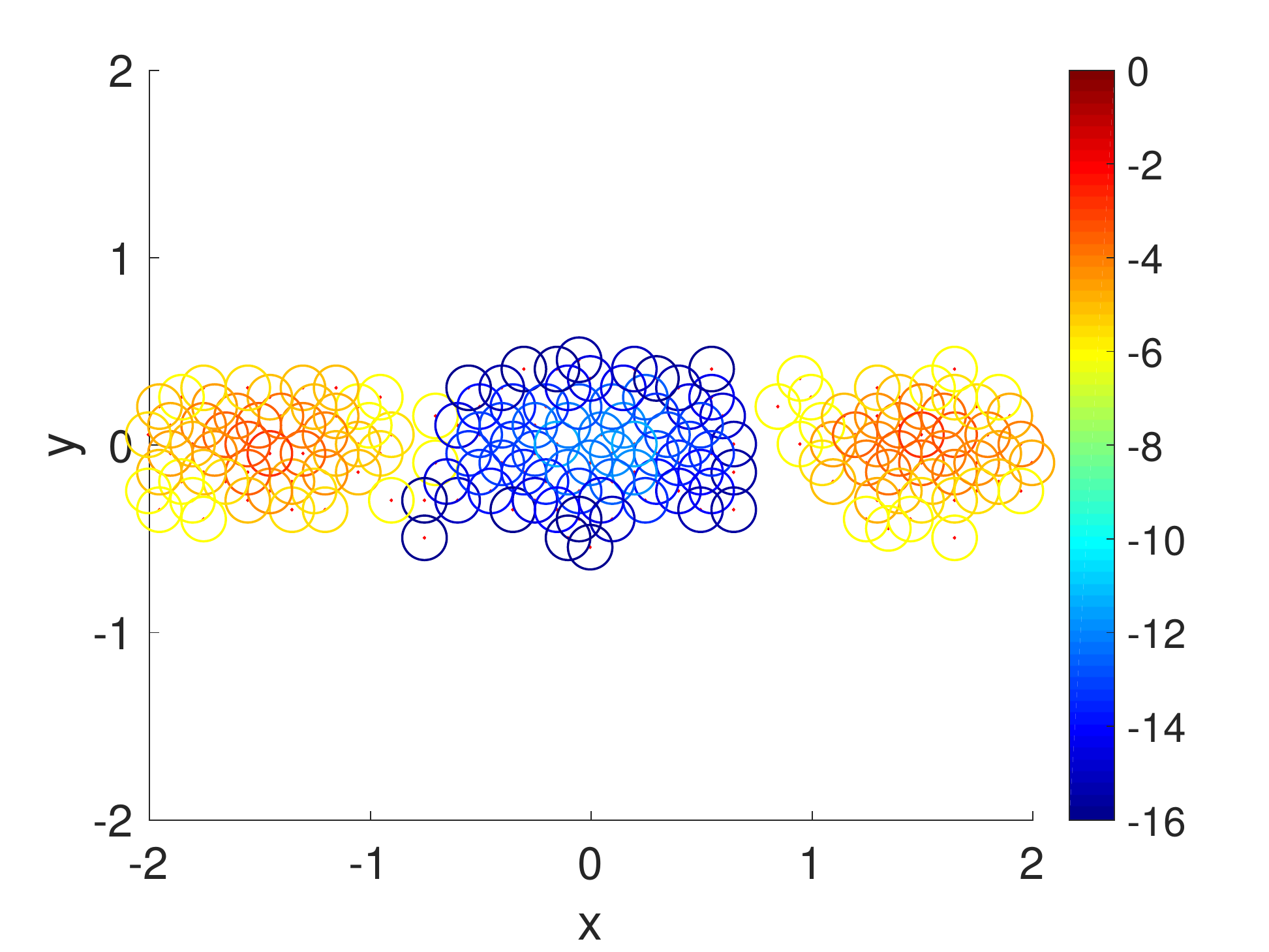} \\
(e) Clusters. Step 133. & (f) After clustering and resampling. Step 133. \\[6pt]
\end{tabular}
\caption{\label{fig:toymodel_sc} Spectral clustering simulation snapshots at $\beta = 25.0$. Voronoi cells are colored according to their weights in log scale in Figures (b), (c), and (f), and the color bar indicates which colors correspond to which weight values in log scale. In Figure (e), each macrostate, a union of Voronoi cells, is represented by a color (five clusters in total, numbered from 0 to 4). For this particular simulation, spectral clustering starts when the number of Voronoi cells is 400 or larger, the total number of walkers per cluster or macrostate $n_{wc}$ is set to 500, and the transition matrix is calculated for 100 steps.}
\end{figure}

To see how spectral clustering works in this example, we plotted simulation snapshots in Fig.~\ref{fig:toymodel_sc}. Spectral clustering or calculation of the transition matrix starts from Step 36 when the total number of Voronoi cells becomes large. After calculation of the transition matrix is finished at Step 136, the committor function is calculated to cluster the Voronoi cells, and the unions of Voronoi cells become the new macrostates. We can see that the macrostates at the very left and righthand sides are the largest, which makes sense since most of the walkers are concentrated near the minimum energy basins $A$ and $B$ and have small weight changes, making them relatively stable. After resampling, however, each macrostates ends up with the same number of walkers, which results in spreading walkers around minimum energy pathway and reducing walkers orthogonal to the pathway. Hence, we can see that the CAS algorithm with spectral clustering is very effective at focusing efforts on sampling slow timescales and is much more efficient than conventional methods, even for this relatively simple two-dimensional model example.

\subsection{\label{sec:penta-alanine} Penta-alanine}
After the low-dimensional model example was tested, we applied the CAS algorithm to high-dimensional real examples. Namely, we applied our method to penta-alanine, which consists of five alanine residues and 66 atoms. Its conformations can be described by the three middle $\phi$ and $\psi$ dihedral angle pairs, or six dihedral angles in total\cite{feng2015}. If all three residues are helical $\alpha_R$ states ($-100.0^{\circ} \leq \phi \leq -30.0^{\circ}$ and $-90.0^{\circ} \leq \psi \leq -10.0^{\circ}$), then penta-alanine is considered to be folded, whereas if all three residues are coiled $C_{7eq}$ states ($-180.0^{\circ} \leq \phi \leq -55.0^{\circ}$ and $105.0^{\circ} \leq \psi \leq 180.0^{\circ}$ or $-180.0^{\circ} \leq \psi \leq -155.0^{\circ}$), then penta-alanine is considered to be unfolded. Since its conformations have to be described in high dimensions, penta-alanine is a nontrivial example. The rates between the two states, or the folding and unfolding rates, are of interest to us. Again, these rates are measured by labeling walkers from folded and unfolded states with colors, changing colors when walkers from one state reach another, and resampling each color within each macrostate, as done in Section~\ref{sec:toymodel}.

The MD simulations are run with Gromacs 4.5.7 using Amber96 force field and implicit solvent at temperature $T = 300\ \mbox{K}$ with time step $\Delta\mbox{t} = 2\ \mbox{fs}$\cite{pronk2013, wang2004}. For brute force simulations, five different initial configurations (folded, unfolded, partially folded, partially unfolded, neither folded nor unfolded) are used, and the five simulations are each run for $3\ \mu\mbox{s}$. The standard deviation of the five runs multiplied by 2, which approximately represent 95\% confidence interval, is used for error bars. The unfolding and folding rates are listed in Table~\ref{tab:bf_table}. For the CAS algorithm simulations, the same MD simulation conditions are used and since all of the collective variables are dihedral angles with $[-180^{\circ},180^{\circ}]$ limits, the minimum distance is taken between the previous and the new dihedral angle values when creating macrostates for the walkers.

\begin{table}
\caption{\label{tab:bf_table} Penta-alanine rates from brute force simulations at $T = 300\ \mbox{K}$.}
\begin{ruledtabular}
\begin{tabular}{ccc}
& Unfolding (folded to unfolded) & Folding (unfolded to folded)\\
\hline
Rate (ns$^{-1}$) & 0.0320$\pm$ 0.0035 & 0.0473$\pm$ 0.0053\\
Error (\%) & 10.79 & 11.24\\
\end{tabular}
\end{ruledtabular}
\end{table}

First, we evaluated the effectiveness of the CAS algorithm compared to conventional brute force simulations. In order to do so, we calculated and plotted the unfolding and folding rates to compare. We use the bootstrapping procedure, or drawing first passage times randomly with replacement for some number of times proportional to the total simulation time, to obtain the rates and to simulate the statistics we would obtain with shorter simulations. For comparison, the CAS algorithm simulations are run with simulation time $\tau = 500.0\ \mbox{ps}$, which happens to fulfill the Markovian property according to Ref.~\onlinecite{feng2015}, even though it is not necessary for the CAS algorithm. Since the rates are not very low, i.e., transitions occur repeatedly due to having low energy barriers, the macrostates are pre-defined and fixed throughout the simulation in this case. The folded and unfolded states are defined as single states, and the rest are partitioned using spectral clustering-related techniques, which prove to give the most optimal macrostates. 

To elaborate, a single $3\ \mu\mbox{s}$ brute force simulation is used to sample the collective variable space, which is then covered with Voronoi cells. These Voronoi cells are used to calculate the transition matrix so that we can obtain the committor function. States are sampled every $500.0\ \mbox{ps}$ so that the transition matrix is Markovian, and $r$ is set to $80.0^{\circ}$ so that approximately 200 Voronoi cells are used to cover all of the states, a size that is shown to give an accurate Markov State Model (MSM) or transition matrix according to Ref.~\onlinecite{feng2015}. Similar to spectral clustering, the committor function is used to cluster the neither folded nor unfolded or intermediate states into macrostates with approximately constant committor function values. Specifically, the minimum and the maximum committor function values are obtained and are used to create an interval of committor function values. This interval is evenly divided into pre-defined number of clusters or macrostates, and the intermediate states are binned to their corresponding macrostates according to their committor function values. This way, every part of the committor function will be efficiently sampled throughout the simulation. The pre-defined number of macrostates is set to be large enough so that none of the macrostates will be empty and to have the macrostates represent a narrow range of committor function values so that they are close to being true isocommittor surfaces. As a result, we end up with a single folded state, a single unfolded state, and a number of states that each represents probability going from folded to unfolded. With these macrostates combined with binning and resampling, the walkers are forced to make progress along the folding/unfolding transition pathway as the simulation proceeds. 

With this pre-defined and fixed macrostate setup, we only need to initialize walkers and choose $n_w$ to run the most optimal CAS algorithm simulations. First, set $n_w$ such that it represents numbers of walkers per macrostate per ``color". In this case, the ``color" represents whether the walker comes from the folded or the unfolded state. For instance, if a macrostate has walkers from both the folded and the unfolded states, then after resampling, the macrostate will end up with $n_w$ walkers that come from the folded state and $n_w$ walkers that come from the unfolded state. Empirically, setting $n_w$ to be the average number of walkers that initially come from the folded state is found to yield the most accurate unfolding rate and vice versa for the folding rate. Hence, $n_w$ is chosen to be 90 for the unfolding rate and 230 for the folding rate. Finally to initialize walkers, all of the states need to make a transition from their initial states to their next states according to what their next states are from the brute force trajectory and resampled according to the pre-defined $n_w$ as previously stated. The resulting walkers are used as initial walkers for the CAS algorithm simulations. 

\begin{figure}
\centering
\begin{tabular}{cc}
\includegraphics[width=80mm]{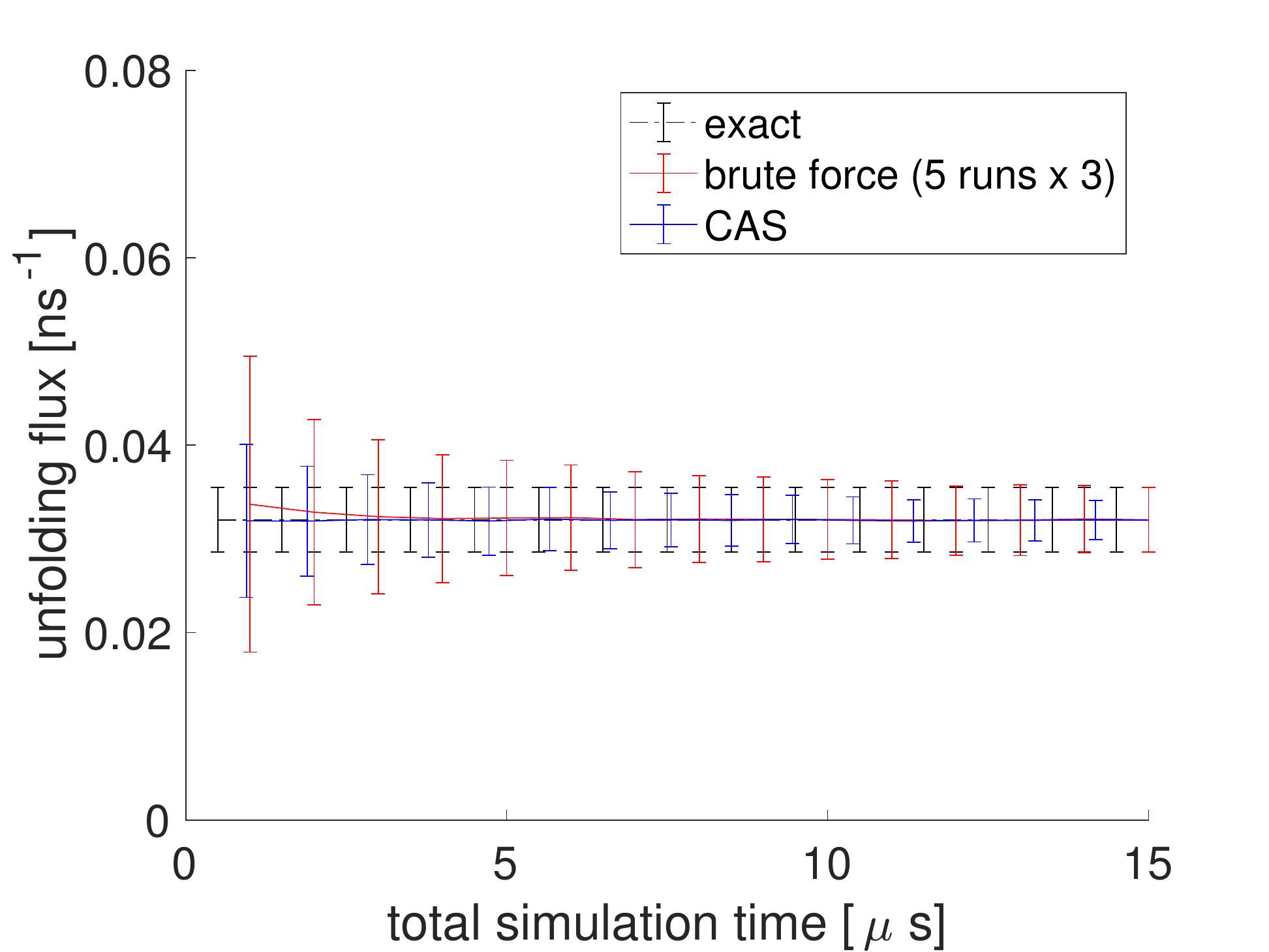} & \includegraphics[width=80mm]{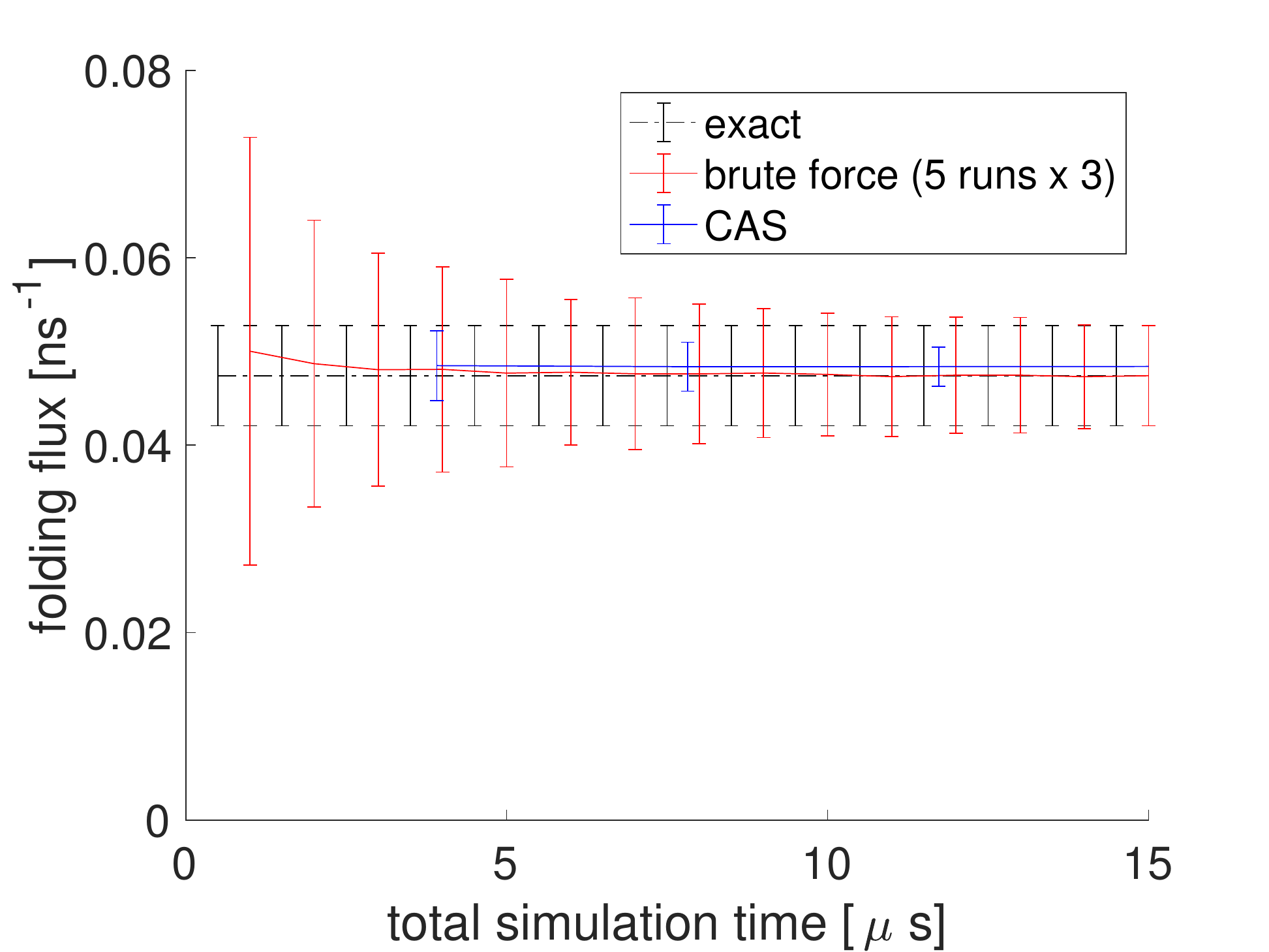} \\
(a) Unfolding rate. & (b) Folding rate. \\[6pt]
\includegraphics[width=80mm]{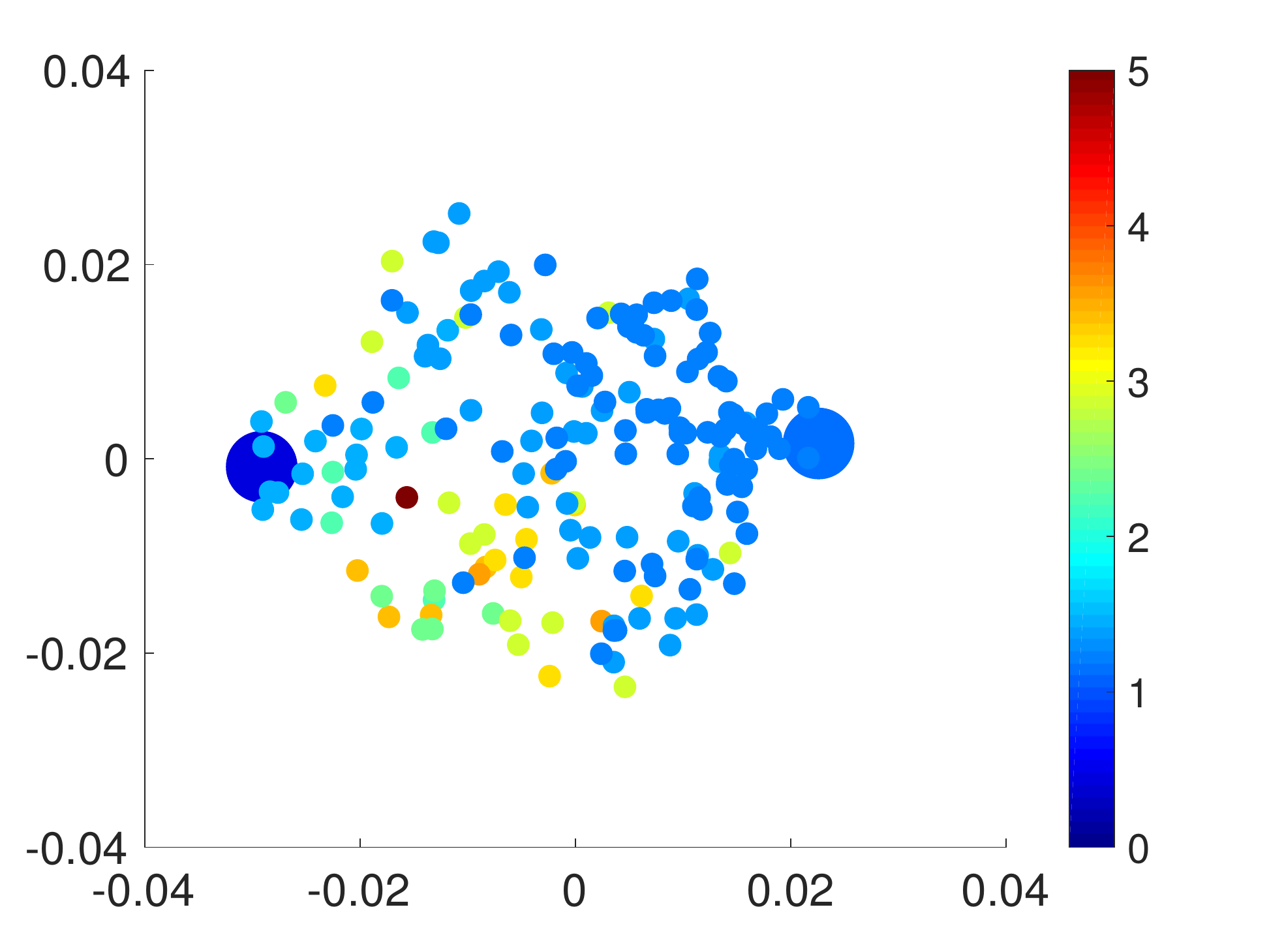} & \includegraphics[width=80mm]{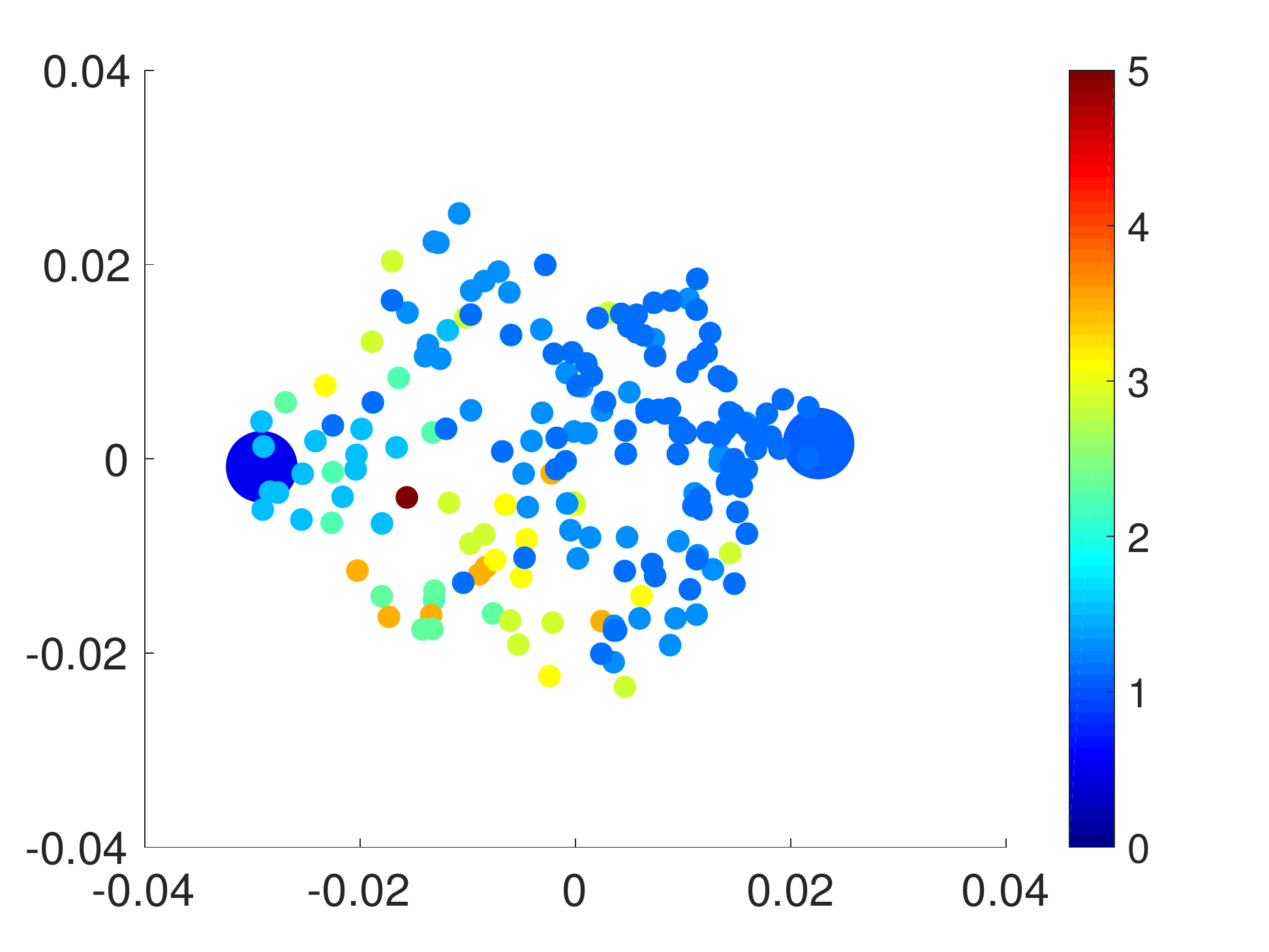} \\
(c) Free energy landscape from brute force. & (d) Free energy landscape from CAS. \\[6pt]
\end{tabular}
\caption{\label{fig:a5_fluxes_plots} Rate and free energy landscape comparisons between brute force and the CAS algorithm simulations at $T = 300\ \mbox{K}$. The macrostates are colored according to their weights in log scale in Figures (c) and (d), and the color bar indicates which colors correspond to which weight values in log scale. The larger circular macrostates on the left and righthand side represent the folded state and the unfolded state, respectively. The $\epsilon$ for diffusion map, which is used for visualization purposes only, is set to 10, and the number of macrostates (excluding the folded and unfolded states) is set to 10.}
\end{figure}

To directly compare the CAS algorithm's accuracy and efficiency with brute force simulations, the CAS algorithm simulations are run for $15\ \mu\mbox{s}$, which is equal to the number of macrostates $\times$ cumulative total number of walkers $\times$ simulation time ($\tau = 500.0\ \mbox{ps}$). As seen in Fig.~\ref{fig:a5_fluxes_plots} and in Table~\ref{tab:cas_table}, the performance of the CAS algorithm is significantly better than brute force in getting the correct rates efficiently with much smaller error bars. 

\begin{table}
\caption{\label{tab:cas_table} Penta-alanine rates from the CAS algorithm simulations at $T = 300\ \mbox{K}$.}
\begin{ruledtabular}
\begin{tabular}{ccc}
& Unfolding (folded to unfolded) & Folding (unfolded to folded)\\
\hline
Rate (ns$^{-1}$) & 0.0320$\pm$ 0.0020 & 0.0484$\pm$ 0.0018\\
Error (\%) & 6.15 & 3.72\\
Reduction in error ($\frac{\text{Brute force error}}{\text{CAS error}}$) & 1.75 & 3.02\\
\end{tabular}
\end{ruledtabular}
\end{table}

Second, we plotted the free energy landscape obtained from brute force and the CAS algorithm to further validate the CAS algorithm's accuracy. To visualize a high-dimensional bio-molecule such as penta-alanine, we used diffusion maps to project the high-dimensional space onto two dimensions\cite{coifman2006, delaporte2008}. Diffusion map is a non-linear dimensionality reduction technique that discovers the underlying low-dimensional manifold, preserves the true geometric structure, and is robust to noise perturbation\cite{delaporte2008}. The parameter $\epsilon$ for diffusion map corresponds to neighbor size and is chosen so that the underlying manifold is clearly shown and not entirely uniformly distributed from one another. For both brute force and the CAS algorithm, $15\ \mu\mbox{s}$ of simulation data is used to plot the weights. As mentioned previously, the folded state and the unfolded state are fixed to be single states and the rest are clustered according to their committor function values. Hence as seen in Fig.~\ref{fig:a5_fluxes_plots}, the folded state (left) and the unfolded state (right) are represented as larger circular macrostates, and the rest are clustered and colored according to their macrostates' weights. The free energy landscape from brute force and the CAS algorithm are almost identical, which validates the sampling accuracy of the algorithm.

\begin{figure}
\centering
\begin{tabular}{cc}
\includegraphics[width=80mm]{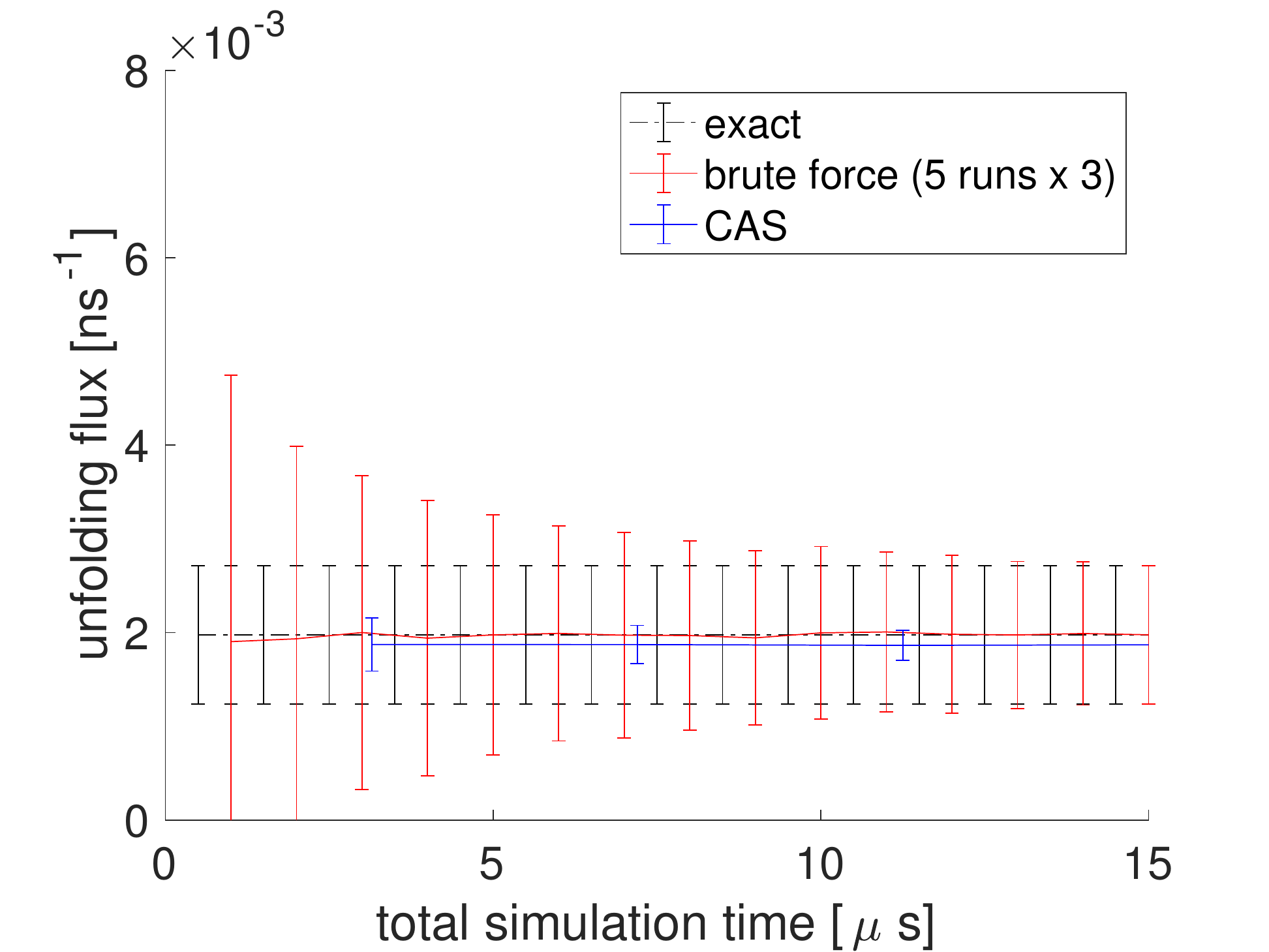} & \includegraphics[width=80mm]{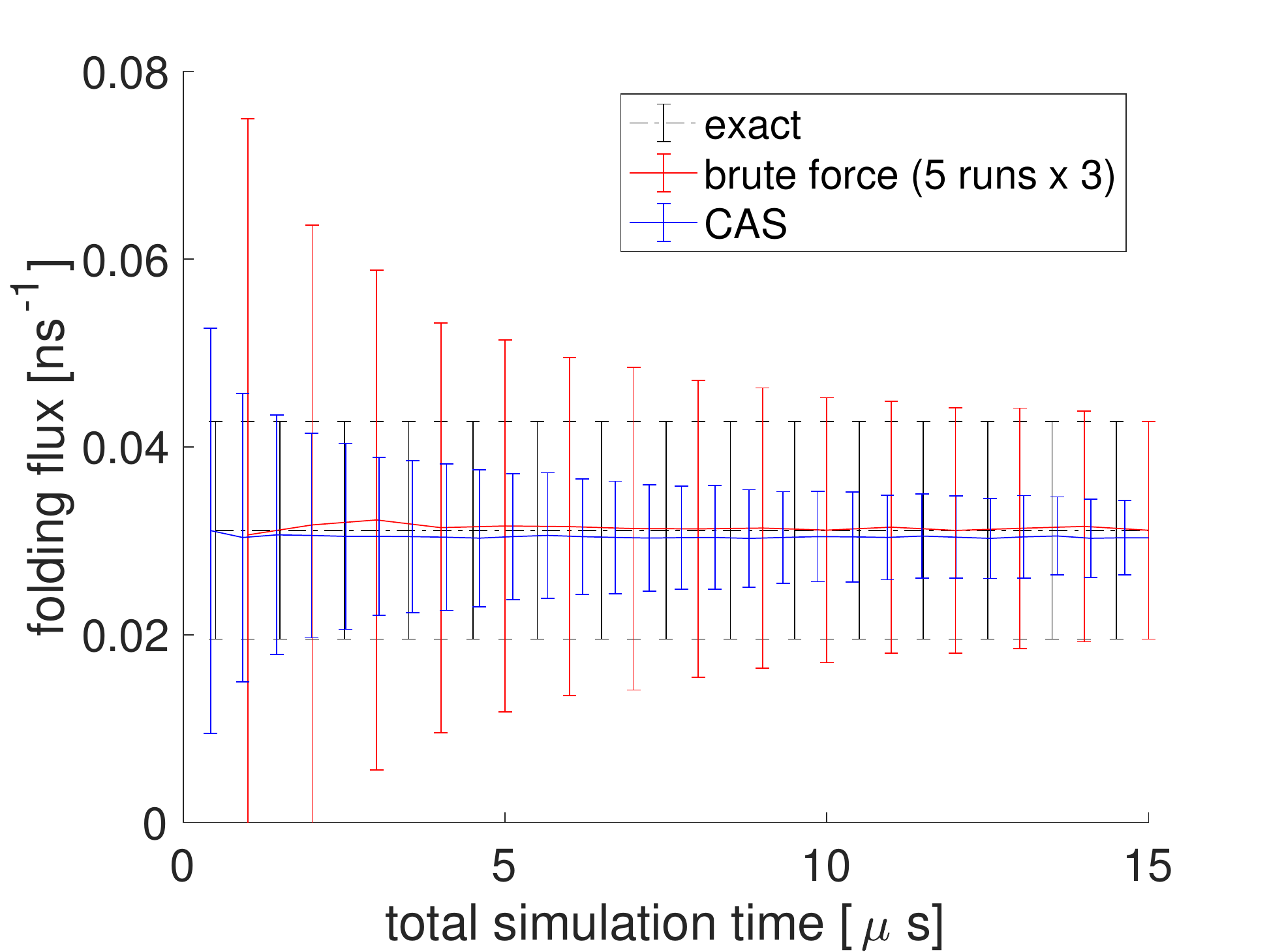} \\
(a) Unfolding rate. & (b) Folding rate. \\[6pt]
\end{tabular}
\caption{\label{fig:a5_fluxes_plots_low_temp} Rate comparisons between brute force and the CAS algorithm simulations at $T = 250\ \mbox{K}$.}
\end{figure}

To see more of CAS algorithm's efficiency over brute force simulations, we lowered the system temperature from $T = 300\ \mbox{K}$ to $T = 250\ \mbox{K}$, which made the transitions significantly rarer. Again, $15\ \mu\mbox{s}$ of brute force simulations are done for the brute force fluxes and a $3\ \mu\mbox{s}$ brute force simulation trajectory is used to partition the free energy landscape beforehand with the committor function for the CAS algorithm. The unfolding and folding rates are listed in Table~\ref{tab:bf_table_250}. For the CAS algorithm simulations at $T = 250\ \mbox{K}$, $n_w$ is chosen to be 150 for the unfolding flux and 20 for the folding flux. As seen in Fig.~\ref{fig:a5_fluxes_plots_low_temp} and Table~\ref{tab:cas_table_250}, the reduction in error is much greater at lower temperatures. 

\begin{table}
\caption{\label{tab:bf_table_250} Penta-alanine rates from brute force simulations at $T = 250\ \mbox{K}$.}
\begin{ruledtabular}
\begin{tabular}{ccc}
& Unfolding (folded to unfolded) & Folding (unfolded to folded)\\
\hline
Rate (ns$^{-1}$) & 0.00198$\pm$ 0.000739 & 0.0311$\pm$ 0.0116\\
Error (\%) & 37.36 & 37.27\\
\end{tabular}
\end{ruledtabular}
\end{table}

\begin{table}
\caption{\label{tab:cas_table_250} Penta-alanine rates from the CAS algorithm simulations at $T = 250\ \mbox{K}$.}
\begin{ruledtabular}
\begin{tabular}{ccc}
& Unfolding (folded to unfolded) & Folding (unfolded to folded)\\
\hline
Rate (ns$^{-1}$) & 0.00187$\pm$ 0.000141 & 0.0303$\pm$ 0.0040\\
Error (\%) & 7.52 & 13.17\\
Reduction in error ($\frac{\text{Brute force error}}{\text{CAS error}}$) & 4.97 & 2.83\\
\end{tabular}
\end{ruledtabular}
\end{table}

\begin{figure}
\centering
\includegraphics[width=0.8\textwidth]{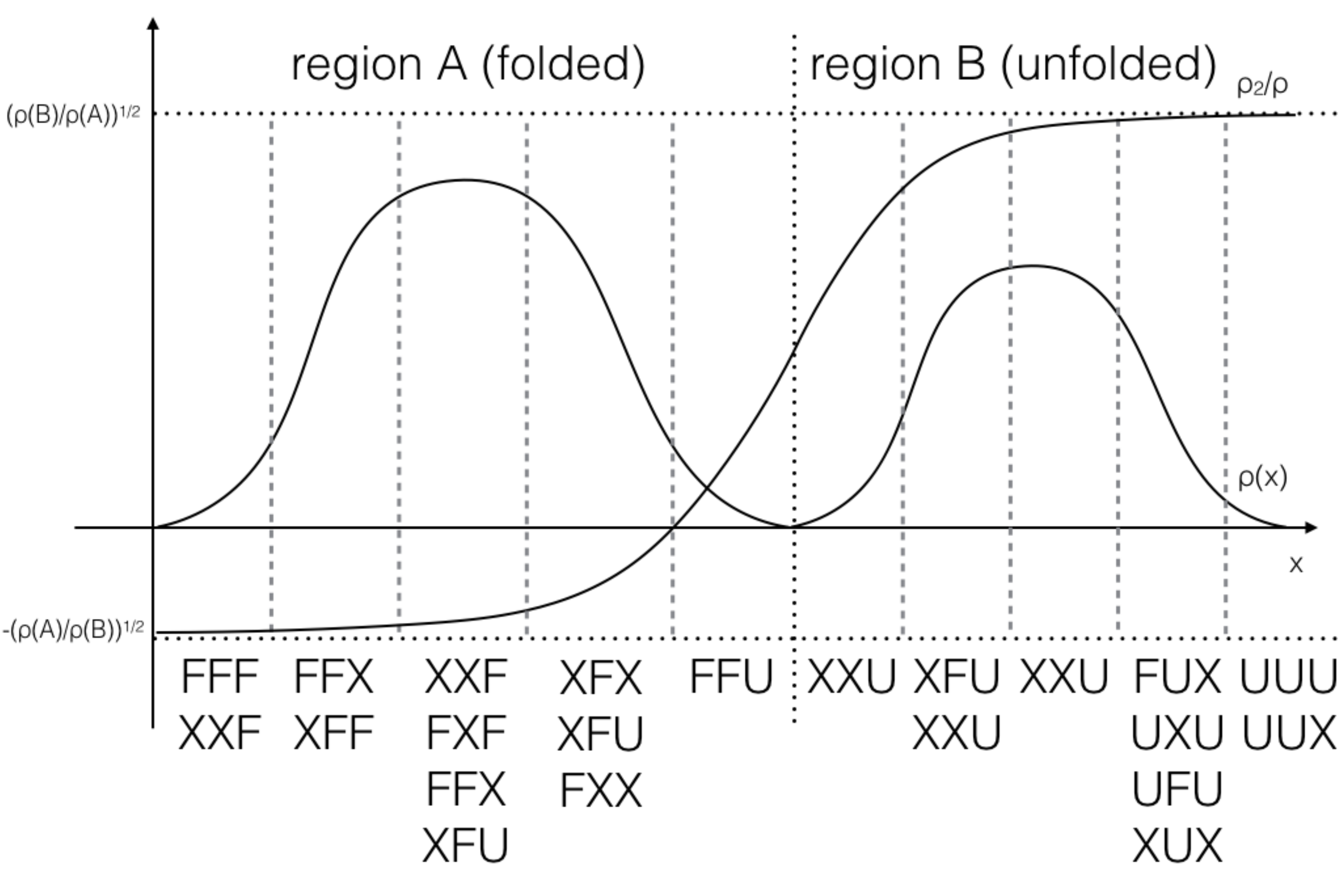}
\caption{\label{fig:a5_committor_confs} Diagram of the committor function and macrostates for penta-alanine at $T = 300\ \mbox{K}$. $\rho(x)$ indicates the equilibrium weights and $\rho_2/\rho$ indicates the committor function. The committor function is uniformly divided into 10 clusters. The folded state (labeled as FFF) and the unfolded state (as UUU) have committor function values that are within the committor values of the leftmost and rightmost macrostates, respectively, so they are marked as the major conformations here, but are separate macrostates from the leftmost and rightmost macrostates.}
\end{figure}

Finally, we extracted the major conformations for each macrostate to check whether they correspond to the correct intermediate states according to their range of committor function values. As expected, the conformations' degree of foldedness/unfoldedness matched with their committor function values and the folded states (labeled as FFF) gradually unfolded one by one as they got nearer the unfolded states (labeled as UUU) and vice versa, as seen in Fig.~\ref{fig:a5_committor_confs}. Interestingly, none of the intermediate macrostates between folded and unfolded states, except for the one nearest the unfolded state, has the first and the second $\phi, \psi$ pairs unfolded. This is consistent with the claim that the first $\phi, \psi$ pair, which is nearest to the N terminus, has the slowest relaxation to unfold, as stated in Ref.~\onlinecite{buchete2008}. Hence, penta-alanine most likely unfolds like a zipper starting from the C terminus and ending with the N terminus.

Taken together, the CAS algorithm is not only able to efficiently obtain kinetic pathways and rates for penta-alanine, but is also able to extract useful thermodynamic information like transition states and free energy landscapes. With the use of diffusion map, we are also able to visualize the high-dimensional conformational space and pathways going from one state to another.

\subsection{\label{sec:triazine_polymers} Triazine polymer}
\begin{figure}
\centering
\begin{tabular}{cc}
\includegraphics[width=60mm]{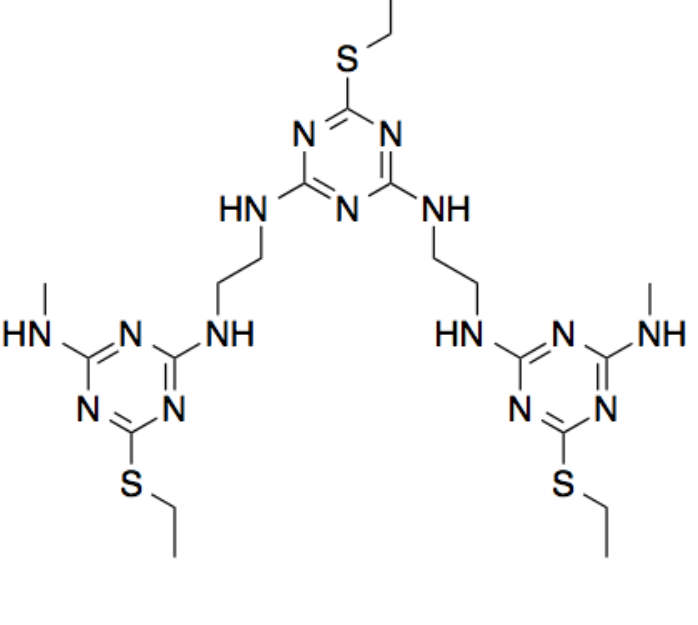} & \includegraphics[width=60mm]{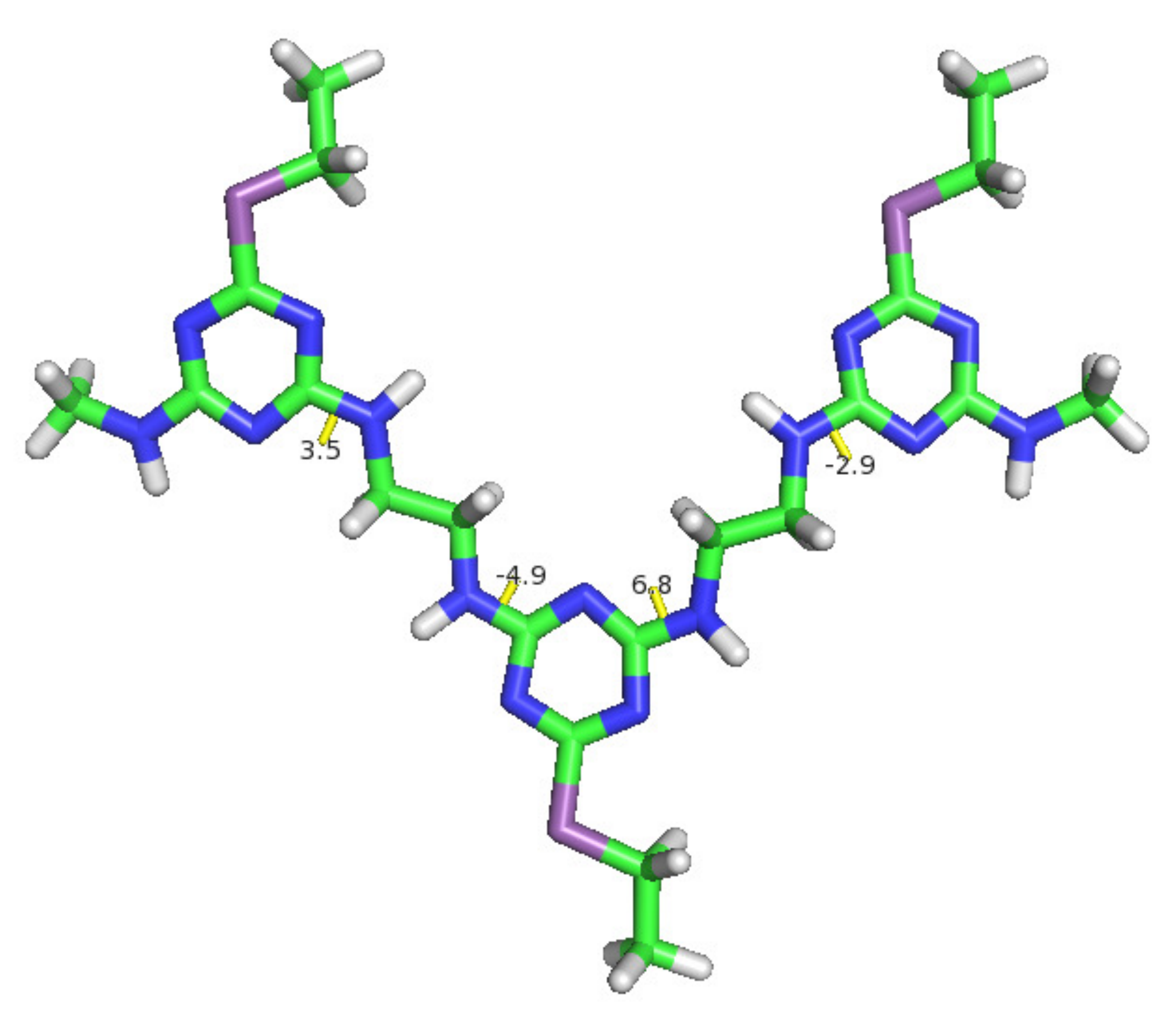} \\
(a) Molecular structure. & (b) Pymol structure with \emph{cis/trans} dihedral angles marked. \\[6pt]
\end{tabular}
\caption{\label{fig:triazine} Structure of \emph{cis}-triazine trimer.}
\end{figure}

For the final example, the CAS algorithm is applied to a high information content triazine polymer newly developed by Grate and others at Pacific Northwest National Laboratory\cite{grate2016}. The triazine polymers encode information by having various side chains and since they do not have hydrolyzable bonds, the molecules are robust and are not susceptible to proteases\cite{grate2016}. Although the triazine polymers have been shown to form particular sequential stacks, have stable backbone-backbone interactions through hydrogen bonding and pi-pi interactions, and conserve the \emph{cis/trans} conformation throughout the simulation, there are still many questions left to be solved. We do not know its various possible conformations along with their probabilities of occurring and the rare pathways and probabilities of \emph{cis}-to-\emph{trans} transitions. Fig.~\ref{fig:triazine} shows the structure of a single \emph{cis}-triazine trimer.

To demonstrate that the CAS algorithm can capture rare pathways not easily accessible by regular MD simulations, the CAS algorithm is used to sample \emph{cis}-to-\emph{trans} transitions. For this, a single \emph{cis}-triazine trimer is simulated with Gromacs 4.6.4 using implicit solvent at temperature $T = 300\ \mbox{K}$ with time step $\Delta\mbox{t} = 2\ \mbox{fs}$\cite{pronk2013}. Most simulation parameters are identical to the ones in Ref.~\onlinecite{grate2016}, including the force field that was generated using the generalized Amber force field (GAFF) and is used with a Generalized Born/Surface Area (GBSA) implicit solvation model\cite{wang2004, voelz2011}. Otherwise, the radius $r$ is set to $24^{\circ}$, the target number of walkers per Voronoi cell $n_w$ is set to 10, and the simulation time $\tau$ is set to $0.04\ \mbox{ps}$. The collective variables are the four dihedral angles that determine the \emph{cis/trans} configuration, which are marked in Fig.~\ref{fig:triazine}. Note that these are not exactly the same as the conventional $\omega$ dihedral angles, which determine the \emph{cis/trans} configuration in peptide bonds. But like the regular $\omega$ dihedral angles, the molecule is cis when the dihedral angles are all equal to $0^{\circ}$ and trans when they are all equal to $180^{\circ}$. Again, as in Section~\ref{sec:penta-alanine}, since all of the collective variables are dihedral angles with $[-180^{\circ},180^{\circ}]$ limits, the minimum distance is taken between the previous and the new dihedral angle values when creating Voronoi cells for the walkers.

\begin{figure}
\centering
\begin{tabular}{cc}
\includegraphics[width=65mm]{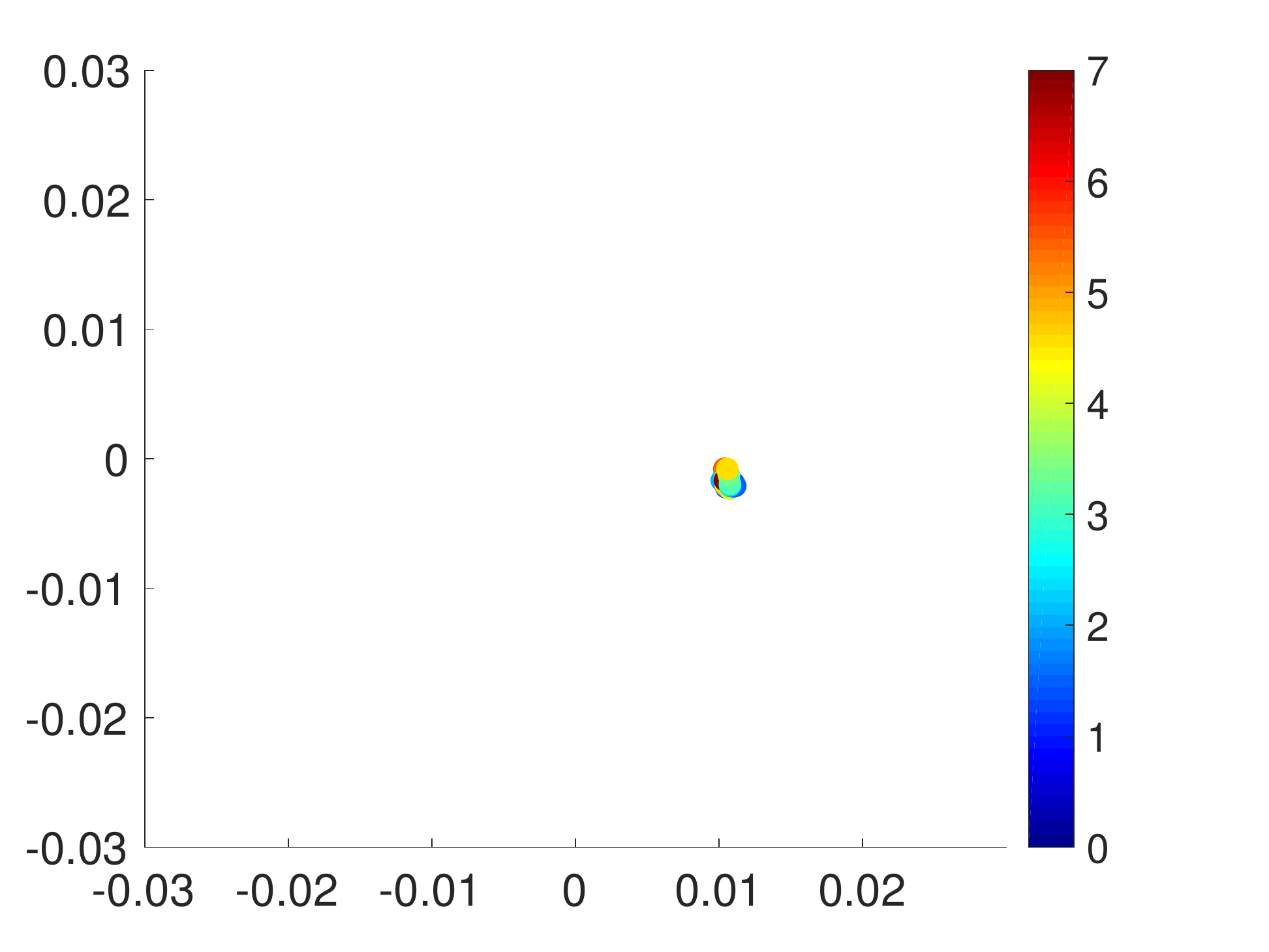} & \includegraphics[width=65mm]{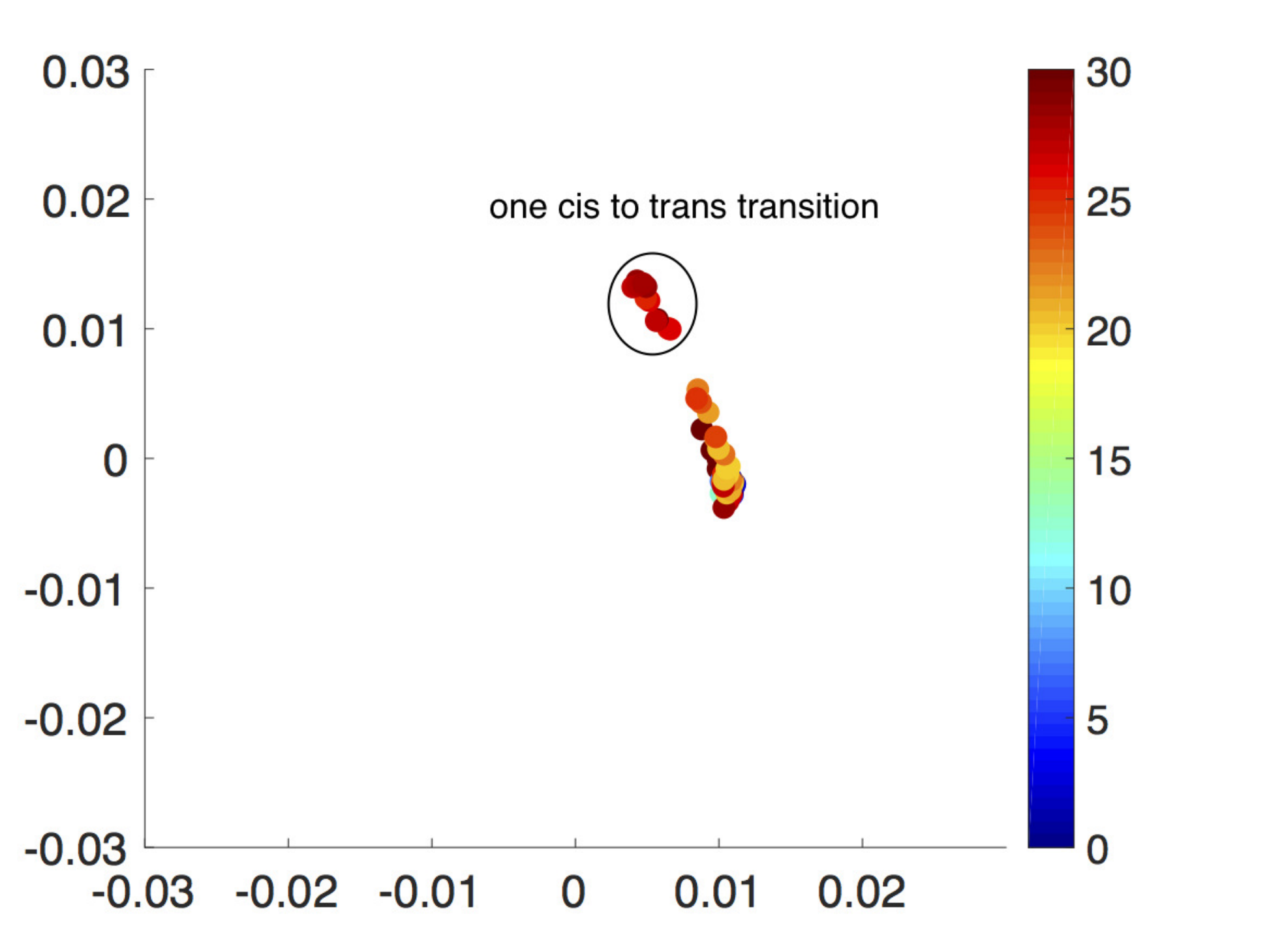} \\
(a) Step 50. & (b) Step 100. \\[6pt]
\includegraphics[width=65mm]{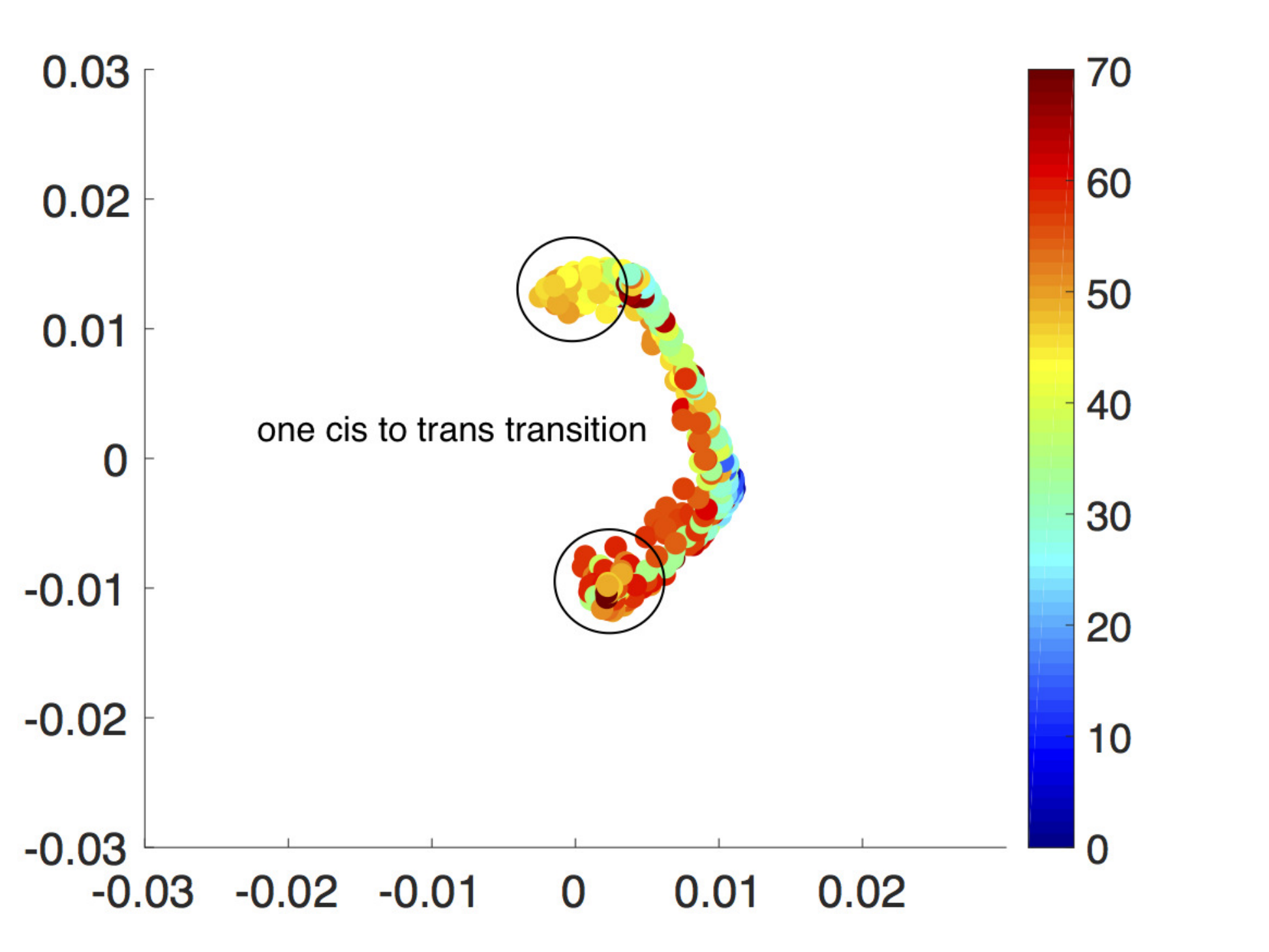} & \includegraphics[width=65mm]{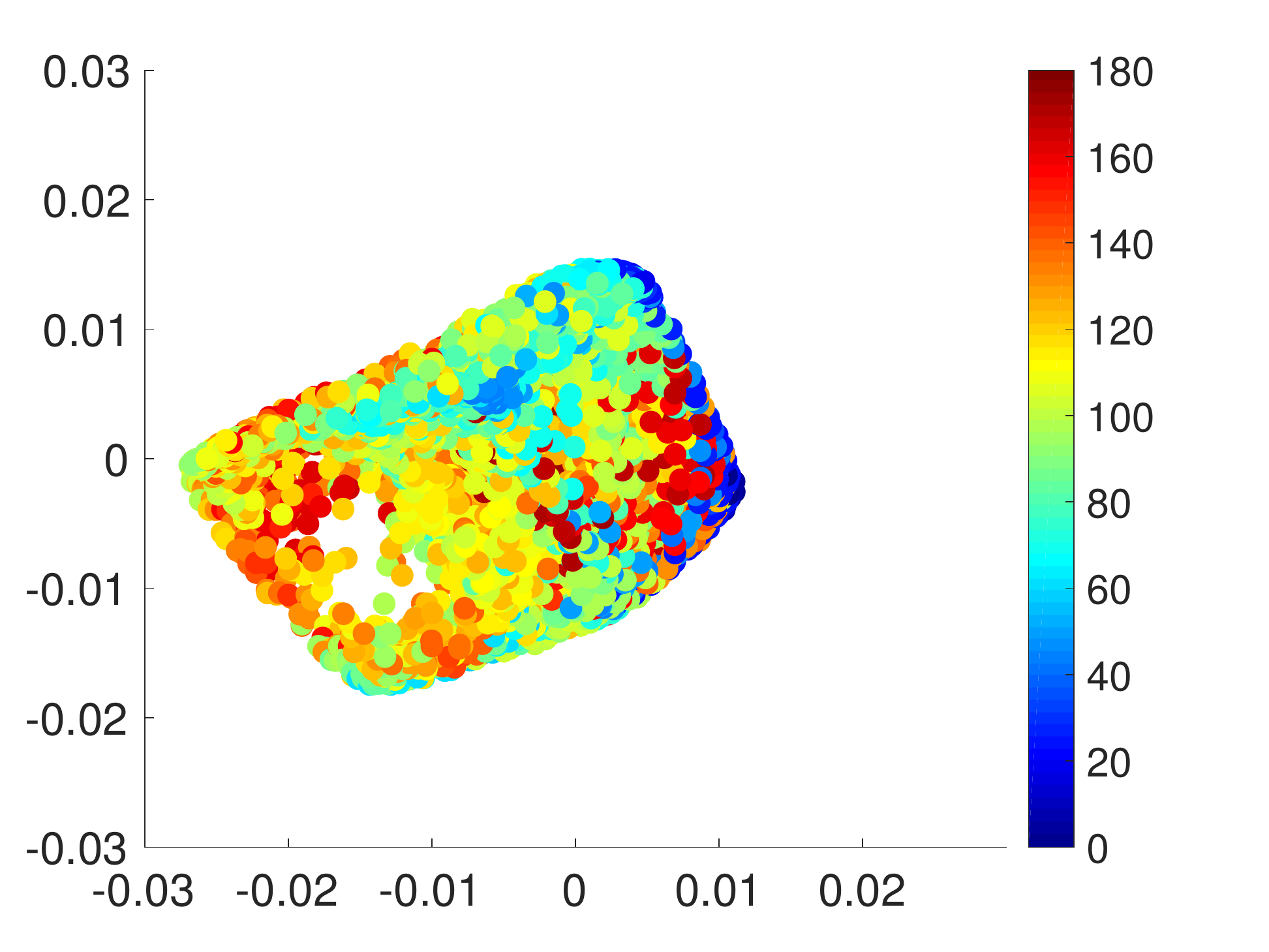} \\
(c) Step 200. & (d) Step 400. \\[6pt]
\multicolumn{2}{c}{\includegraphics[width=65mm]{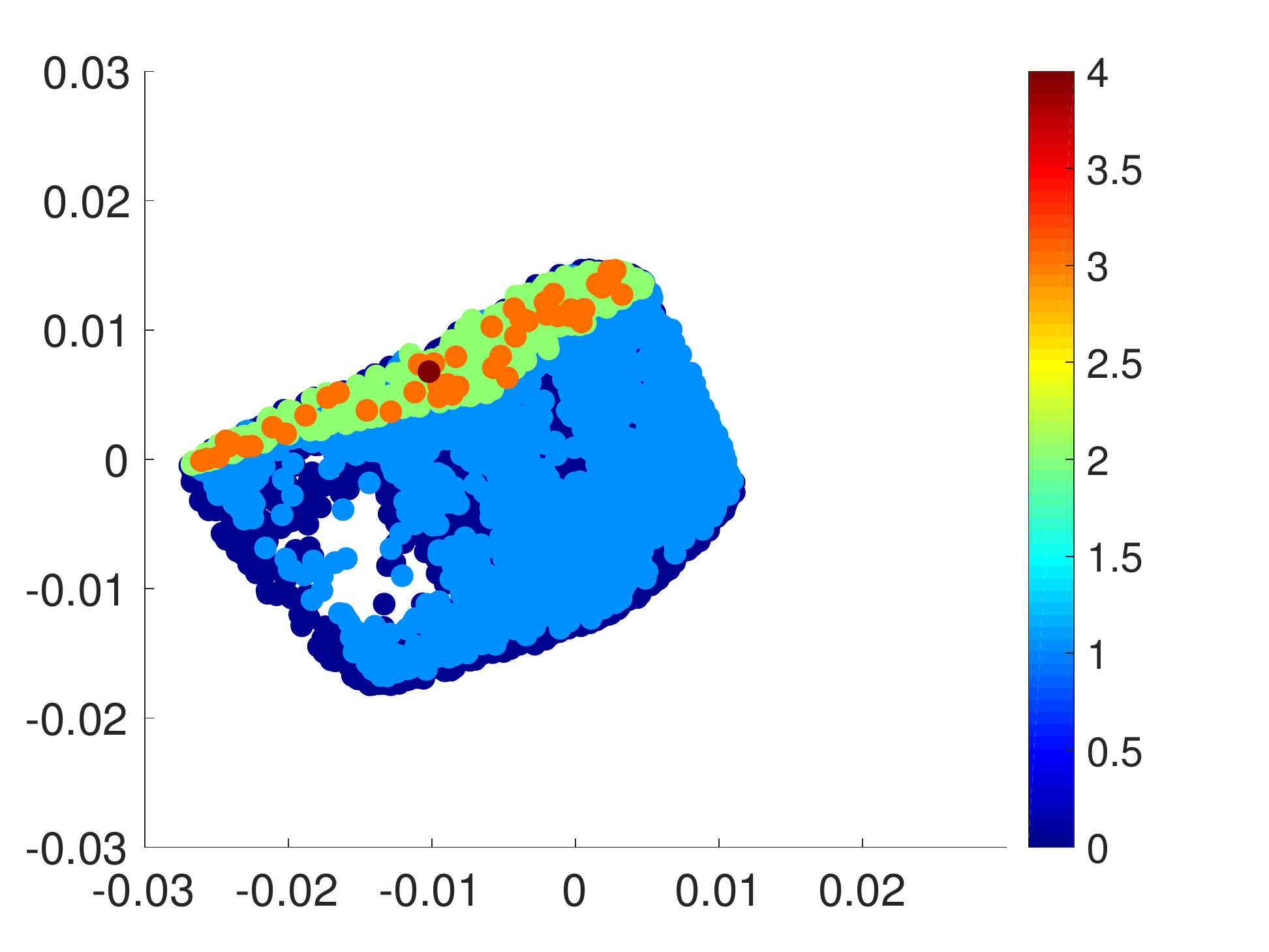}}\\
\multicolumn{2}{c}{(e) Number of \emph{cis} to \emph{trans} transitions.}
\end{tabular}
\caption{\label{fig:triazine_cis_to_trans} Simulation snapshots of \emph{cis}-triazine trimer transitioning to \emph{trans}-triazine trimer. The Voronoi cells are colored according to their weights in log scale and the color bar indicates which colors correspond to which weight values in log scale, except for Figure (e) where the color bar indicates the number of \emph{cis} to \emph{trans} transitions. The $\epsilon$ for diffusion map, which is used for visualization purposes only, is set to 1.}
\end{figure}

Since the energy barrier to get to \emph{trans} configuration is very high, the walkers tend to cluster around the \emph{cis} region initially. However, with a very short resampling time, small Voronoi cells, and enough number of walkers per Voronoi cell, the all \emph{cis}-triazine trimer is able to go from \emph{cis} to \emph{trans} one dihedral angle at a time and eventually, it transitions into an all \emph{trans}-triazine trimer as seen in Fig.~\ref{fig:triazine_cis_to_trans}. Again, since this is a four-dimensional problem, diffusion map is used to visualize each step of the CAS algorithm simulation. 

From running the CAS algorithm simulation for 700 resampling steps, the transition from \emph{cis}-triazine trimer to \emph{trans}-triazine trimer is calculated to have a probability of $9.22\times 10^{-60}$ or an energy barrier about $80\ \mbox{kcal/mol}$, assuming $T = 300.0\ \mbox{K}$, which makes sense since there are four \emph{cis}-to-\emph{trans} transitions and each one has a probability of $2.69\times 10^{-15}$ or an energy barrier about $20\ \mbox{kcal/mol}$\cite{craveur2013}. This is a remarkable result, since the CAS algorithm is able to observe an event with infinitesimal probability, which would take $4\times 0.04\ \mbox{ps}\ \times 1/2.69\times 10^{-15} \approx 1\ \mbox{minute}$ in real time and $2.97\times 10^{16}$ simulation time steps assuming $\Delta\mbox{t} = 2\ \mbox{fs}$. Needless to say, the \emph{cis}-to-\emph{trans} transitions have not been observed in conventional MD simulations since the energy barrier is too high and with the CAS algorithm, we were easily able to observe these rare transitions with optimal choice of parameters. 

\section{\label{sec:discussion} Discussion}
The CAS algorithm is an efficient simulation method that combines a method of enhanced sampling to overcome energy barriers and a method based on the transition matrix to reduce redundant walkers that do not allow the simulation to efficiently make progress along the slowest reaction. The novel and important features of the method are the following:

\begin{enumerate}
\item There is no Markovian error as in MSMs, which requires a global convergence of the weights. This is a necessary trade-off, however, resulting from the fact that Markovian approximation is not used and therefore, we need to wait until steady-state is reached, at which point the walkers are correctly distributed and non-Markovian effects disappear (Chapter 7 in Schlick\cite{schlick2012}).
\item Exact rates are obtained upon convergence with no bias.
\item Optimal macrostates are constructed, which have small statistical errors.
\item Computational cost is strictly controlled by reducing the aforementioned redundant walkers, while allowing the simulation to make progress in sampling the slowest pathway.
\item Mild assumptions or little a priori knowledge about the system is required, since the partitioning relies on Voronoi cells. This is useful for relatively unknown or unfamiliar systems like the triazine polymers in Section~\ref{sec:triazine_polymers}.
\item General collective variables, including non-differential variables such as discrete coordinates with integer or boolean values, can be considered. This will be useful for further studying the triazine polymers' self-assembly (e.g., the number of hydrogen bonds and pi-pi interactions as a collective variable).
\item The MD simulation program can be used in a black box manner, i.e., the wrapper Python CAS algorithm code can be used with any MD simulation program without having to modify its source code and is available at \url{http://github.com/shirleyahn/CAS_Code}.
\item There is a large amount of parallelism in the algorithm, since we simultaneously run many walkers for each macrostate. This allows us to achieve computational efficiency proportional to available computational resources.
\end{enumerate}

Furthermore, the CAS algorithm focuses on identifying critical pathways and transition states and is able to extract thermodynamic and kinetic information in a general setting. The CAS algorithm is also not hampered by presence of metastable states since a constant stream of walkers at visited macrostates is maintained by resampling. Finally, we only need to tune a few parameters to increase efficiency in the sampling, such as simulation time, number of walkers per macrostate, and macrostate size. 

\begin{acknowledgments}
This work is supported by the Applied Mathematics Program within the Department of Energy (DOE) Office of Advanced Scientific Computing Research (ASCR) as part of the Collaboratory on Mathematics for Mesoscopic Modeling of Materials (CM4). We thank Hee Sun Lee for contributing to the regular WE method code that served as the basis for the CAS algorithm code and Johannes Birgmeier for making improvements on the parameter input file. 
\end{acknowledgments}

\bibliography{CAS_Algorithm}

\end{document}